\documentclass[11pt]{article}
\pdfoutput=1  

\usepackage[T1]{fontenc}
\usepackage[utf8]{inputenc}
\usepackage{newtxtext}                       
\usepackage[a4paper,top=1in,bottom=1in,left=1.25in,right=1.25in]{geometry}
\usepackage{amsmath}
\usepackage{amssymb}
\usepackage{newtxmath}                        
\usepackage[scaled=0.92,varqu]{zi4}           
\usepackage{booktabs}
\usepackage{longtable}
\usepackage{array}
\usepackage{graphicx}
\usepackage{microtype}
\usepackage[table,svgnames]{xcolor}
\usepackage{caption}
\definecolor{rowgray}{gray}{0.94}
\definecolor{headcol}{HTML}{E7EEF3}
\usepackage[round,sort&compress]{natbib}
\definecolor{linkblue}{HTML}{1A4FD6}
\usepackage{hyperref}
\hypersetup{colorlinks=true,citecolor=linkblue,linkcolor=linkblue,urlcolor=black}
\usepackage{xurl}
\usepackage{enumitem}
\usepackage{parskip}
\renewenvironment{abstract}{%
  \small
  \begin{center}{\large\bfseries\sffamily\abstractname}\end{center}%
  \begingroup\leftskip1.5em\rightskip1.5em\noindent\ignorespaces}{\par\endgroup}
\usepackage[htt]{hyphenat}
\usepackage{titlesec}
\titleformat*{\section}{\Large\bfseries\sffamily}
\titleformat*{\subsection}{\large\bfseries\sffamily}
\titleformat*{\subsubsection}{\normalsize\bfseries\sffamily}
\titleformat*{\paragraph}{\normalsize\bfseries\sffamily}

\newcommand{\nDCG}{nDCG@10}

\newcommand{\model}[1]{\texttt{#1}}
\newcommand{\secref}[1]{\hyperref[#1]{\S\ref*{#1}}}
\newcommand{\appref}[1]{\hyperref[#1]{Appendix~\ref*{#1}}}
\graphicspath{{figures/}}

\title{\large\sffamily\bfseries Bekko Embedding: Parameter-Efficient Multilingual Retrieval\\ with Ultra-Compact Encoders}
\author{Yuichi Tateno \textless hotchpotch@gmail.com\textgreater}
\date{}

\begin{document}
\maketitle

\begin{abstract}
Dense retrieval models now commonly run to hundreds of millions or billions of parameters---costly in inference time and memory on CPU-only servers, edge devices, and web browsers---yet few models combine strong retrieval quality, broad multilingual coverage, and a small, fast footprint. We present Bekko Embedding, a family of parameter-efficient multilingual retrieval models built around ultra-compact contextual encoders. Inference compute (FLOPs) is chiefly determined by the parameters that every token passes through---those of the Transformer layers---which we call Active Parameters (AP). Bekko minimizes exactly these: the released models, \model{bekko-embedding-v1-a8m}\footnote{\url{https://huggingface.co/hotchpotch/bekko-embedding-v1-a8m}} and the higher-quality \model{bekko-embedding-v1-a25m}\footnote{\url{https://huggingface.co/hotchpotch/bekko-embedding-v1-a25m}} (below, a8m / a25m), have just under 8M and 25M AP.

The recipe is deliberately simple. We prune the 22 layers of the multilingual encoder mmBERT-small to 4 / 13 layers and train the pruned models as base models in two stages: large-scale contrastive learning on about 1.1 billion multilingual pairs from our public corpus (including two complementary LLM-synthesized datasets), followed by hard-negative fine-tuning. Both stages share one loss that combines a masked contrastive loss with pair-type-dependent loss direction and the Matryoshka Representation Learning objective; no teacher distillation is used, and all training completes on a single GPU---about 3 days for a8m. Despite their size, both models match or surpass models whose AP is one to two orders of magnitude larger. On official MMTEB Multilingual v2 Retrieval (\nDCG{}), a8m scores 56.2, above the multilingual-e5 family and BGE-M3 (40$\times$ the AP) in our comparison; a25m reaches 57.5; and the lightweight Multilingual NanoBEIR (14 languages) confirms the trend. The 384-dimensional output (truncatable to 256/128/64) also keeps downstream similarity search and indexing cheap.

Small AP pays off directly in speed: among the compared models measured on the same workload and hardware, a8m is the fastest on both CPU and GPU---about 1.6$\times$ multilingual-e5-small on x86 CPU and 1.5$\times$ on GPU---and the fastest on a Raspberry Pi 5. Alongside general-purpose PyTorch weights, we ship ONNX / OpenVINO builds whose vocabulary embedding matrix is compressed by row-wise int8 quantization, shrinking the a8m model file to 124~MiB (about one third of fp32) and enabling in-browser execution via Transformers.js. To support reproducible research on efficient multilingual retrieval, we release the model weights, the complete stage-1 corpus, and the independently mined stage-2 hard negatives.

\smallskip
\noindent\textbf{Keywords:} text embeddings; multilingual retrieval; layer pruning; parameter efficiency; on-device inference.
\end{abstract}


\section{Introduction}
\label{sec:intro}

\subsection{The practical value of encoder models}
\label{sec:encoder-value}

Since BERT \citep{devlin2019bert}, encoder-only Transformers have remained central to non-generative applications such as information retrieval (IR), classification, and named entity recognition. The reason encoder models continue to be used in production despite the rapid progress of LLMs lies in their low inference cost and high quality per parameter. LLMs can serve as powerful embedding backbones, but their inference latency and memory footprint often make them impractical for industrial applications \citep{granite2025embedding}. \citet{weller2025ettin}, in a systematic comparison using the Ettin suite of size-matched encoder/decoder pairs, report that encoder-only models outperform decoder-only models on classification and retrieval tasks. In IR in particular, bi-encoder embedding retrieval is indispensable as the core of first-stage retrieval that rapidly narrows a large document collection down to relevant candidates \citep{reimers2019sbert, karpukhin2020dpr}, and as the retrieval backbone of Retrieval-Augmented Generation (RAG) \citep{lewis2020rag}.

\subsection{The right axis of efficiency: Active Parameters, not total parameters}
\label{sec:ap-axis}

The mainstream of recent embedding-model research has moved toward scale: GTE-multilingual-base \citep{zhang2024mgte} uses 305M parameters, BGE-M3 \citep{chen2024bgem3} 568M, and Qwen3 Embedding \citep{zhang2025qwen3embedding} 0.6B--8B (parameter counts are measured from each public model and model card). Against this trend, we first re-examine which axis ``size'' should be measured on.

In multilingual models, the vocabulary embedding matrix (vocabulary $\times$ hidden size) accounts for most of the total parameters. In mmBERT-small \citep{marone2025mmbert} (140M total), for example, $256{,}000 \times 384 \approx 98$M parameters sit in this matrix, and only 42M are non-embedding parameters. Yet at inference time this matrix is a static lookup table indexed by token id: it involves no matrix multiplication, costing only the read-out of one row (hidden-size values) per token, independent of vocabulary size (nor does the output side incur a $\mathrm{vocab} \times \mathrm{hidden}$ projection, since the hidden states are pooled). By contrast, the parameters of the Transformer layers, through which every token must pass, chiefly determine inference compute (FLOPs). In this paper we define Active Parameters (AP) as the learned parameters that the contextual encoder uses at inference time, excluding the embedding table---i.e., the Transformer blocks, normalization, and pooling/projection; note that this definition is ours and differs from the ``active parameters'' of the Mixture-of-Experts literature. Treating non-embedding parameters as the axis of efficiency or scaling is itself established: \citet{kaplan2020scaling} define the parameter count $N$ in their scaling laws as the count excluding vocabulary and positional embeddings, and ALBERT \citep{lan2019albert} factorized the vocabulary embedding away from the hidden dimension to compress it. Our work stands on this axis: we take AP reduction---and hence FLOPs reduction and faster inference---as the primary goal, and additionally shrink the lookup table (vocabulary embedding matrix) by int8 compression to reduce memory and distribution size (\secref{sec:efficiency}). Note that FLOPs also depend on sequence length, attention pattern, and hidden size, and real latency further depends on backend kernels and memory bandwidth, so AP is not the sole determinant of inference cost but its main correlated axis. Since models with similar AP can differ in real speed depending on architecture and inference implementation, we also report measured throughput (\secref{sec:speed}).

Accordingly, this paper adopts AP, rather than total parameters, as the axis of computational efficiency for embedding models. Total-parameter comparisons make small models with large multilingual tokenizers look disproportionately ``big'': multilingual-e5-small \citep{wang2024me5}, for instance, is nominally 118M parameters in total, but its AP excluding the vocabulary embedding is only about 21M. We apply this axis consistently and show that (i) a model with minimized AP can rival models one to two orders of magnitude larger (\secref{sec:results}), and (ii) for edge and low-spec distribution, int8-quantizing the static vocabulary embedding table---the bulk of the distribution size---greatly reduces distribution size and load-time memory without touching the Transformer computation (\secref{sec:efficiency}).

A second efficiency axis is the output embedding dimensionality. Beyond model inference cost, the post-embedding search stage grows linearly in output dimension, in both vector similarity computation and index memory. Whereas recent large models emit 1024--4096-dimensional outputs, Bekko's standard output is only 384-dimensional (further truncatable to 256/128/64 via Matryoshka; \secref{sec:recipe}), which directly reduces search and indexing costs over large document collections.

\begin{figure}[t]
  \centering
  \includegraphics[width=0.92\linewidth]{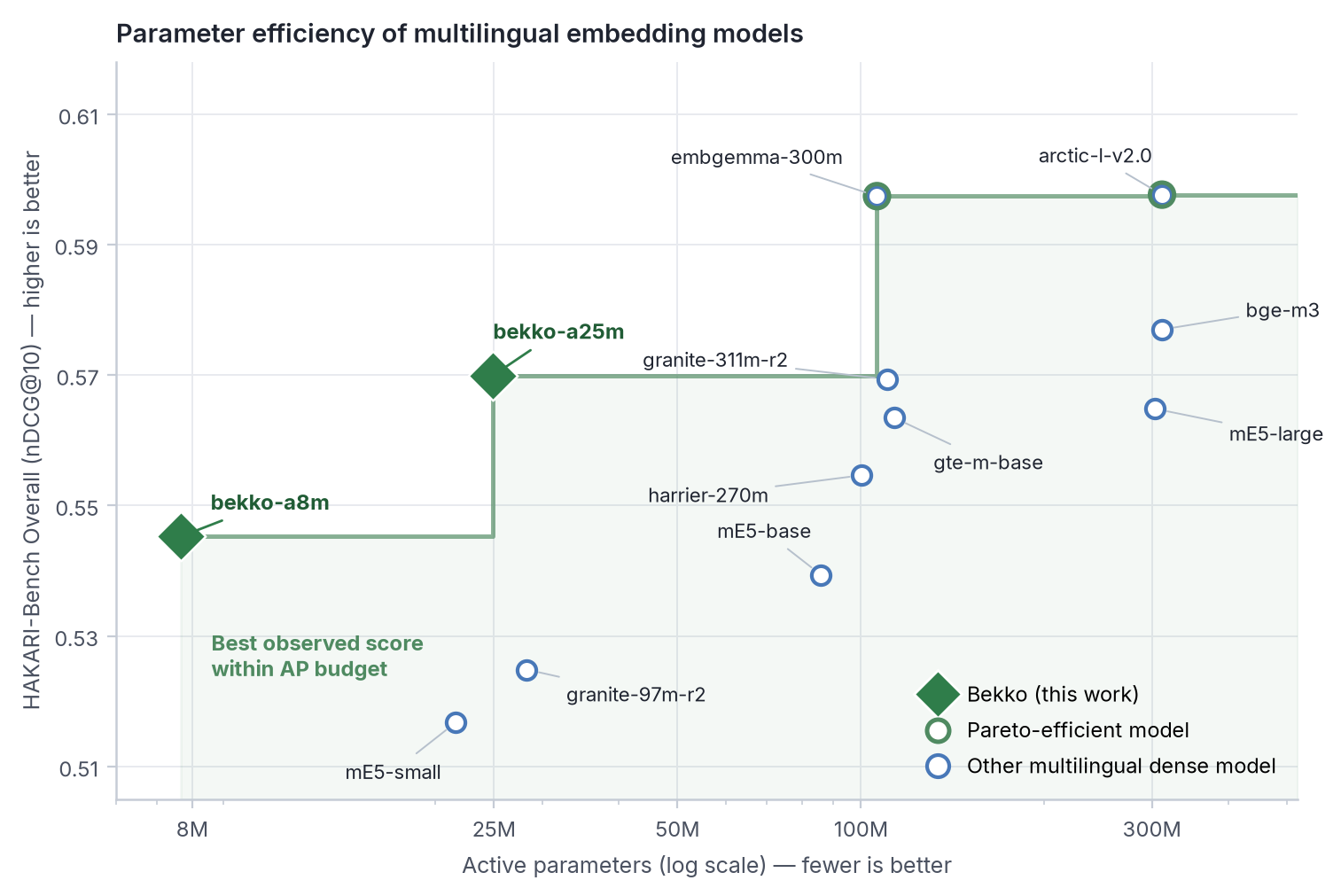}
  \caption{Parameter efficiency of multilingual embedding models: retrieval quality (HAKARI-Bench Overall, a lightweight benchmark created by the author; \secref{sec:evalbench}) versus Active Parameters (log scale). The staircase shows the best observed score attainable under a given AP budget; it does not interpolate between measured models. a8m and a25m extend the efficiency frontier of the unified comparison set (\secref{sec:models}) into a band one to two orders of magnitude smaller in AP. The same picture is confirmed on the primary evaluation, official MMTEB Multilingual v2 (\secref{sec:mmteb}, Figure~\ref{fig:pareto}).}
  \label{fig:intro}
\end{figure}

Figure~\ref{fig:intro} previews Bekko's position on the AP axis. In the plane of retrieval quality (the overall score of the multilingual IR benchmark HAKARI-Bench; \secref{sec:evalbench}) versus AP, a8m and a25m sit on the efficiency frontier of the unified comparison set (\secref{sec:models}) and extend that frontier into the ultra-compact band of 8M--25M AP---achieving retrieval quality on par with models whose AP is one to two orders of magnitude larger, at a far smaller inference cost.

\subsection{On-device demand and the absence of modern ultra-compact multilingual models}
\label{sec:ondevice}

Demand for small, fast, locally runnable retrieval models is rising in real-world IR systems from several directions. (1) When embedding millions to billions of documents, inference cost directly determines processing time and index-construction expense. (2) AI agents that run on machines without high-end GPUs---browsers, smartphones, CPU-only servers---have become practical, creating demand for low-latency retrieval that completes on-device without sending data out; inference in these environments runs on limited compute such as CPU / WebAssembly, browser WebGPU, or smartphone accelerators. Indeed, EmbeddingGemma \citep{schechtervera2025embeddinggemma} targets ``low-latency and high-throughput use cases such as on-device applications,'' and the multilingual Granite Embedding R2 \citep{granite2026multilingualr2} positions its 28M-AP small model for ``latency-sensitive production workloads, edge deployment'' (granite-embedding-97m-multilingual-r2 model card\footnote{\url{https://huggingface.co/ibm-granite/granite-embedding-97m-multilingual-r2}}). (3) Embeddings of new (uncached) queries are computed online at search time, so low-latency inference governs perceived search speed \citep{reimers2019sbert}.

The multilingual ultra-compact band, however, has been dominated by older architectures. In the small-AP range ($\leq$30M), models such as Multilingual E5-small \citep{wang2024me5}---initialized from a multilingual MiniLM\footnote{\url{https://huggingface.co/microsoft/Multilingual-MiniLM-L12-H384}} distilled from XLM-R \citep{conneau2020xlmr}---have long been the default; mE5-small is in essence a 2020-generation MiniLM with about 21M AP. Only in 2026 did Granite Embedding R2 (28M AP) \citep{granite2026multilingualr2}, with a ModernBERT-style design, appear. Yet at least within our comparison criteria (multilingual dense embedding models with openly released weights under commercially usable licenses; \secref{sec:models}), no multilingual contextual encoder occupies the still smaller ultra-compact band of AP $\leq$ 10M that this work targets. In the even smaller regime, Model2Vec \citep{tulkens2024model2vec} and Sentence Transformers static embeddings \citep{aarsen2025static} are prominent for CPU speed, but they are static token embeddings without contextualization and occupy a different quality band from contextual encoders.

\subsection{Our approach and contributions}
\label{sec:approach}

We address this gap with a design that integrates modern architectures and training methods into ultra-compact models. Two keys underlie the approach. First, the multilingual knowledge useful for retrieval already resides in the pretrained weights of the base mmBERT-small; structural layer pruning preserves it while removing layers, and using the pruned model as the base model (the initialization for contrastive learning) lets training succeed with AP compressed to single-digit millions, without distillation. Second, the expressiveness a small model lacks can be compensated for by broad-domain, broad-language contrastive data plus high-quality synthesis and hard negatives (the data design and its debts to prior work are detailed in \secref{sec:data}). Our contributions are:

\begin{enumerate}[leftmargin=*]
\item Consistent application of the AP axis: we apply the view of non-embedding parameters as the efficiency axis \citep{kaplan2020scaling, lan2019albert} consistently to multilingual embedding-model comparison, ultra-compaction, int8 compression, and edge distribution (\secref{sec:ap-axis}, \secref{sec:models}, \secref{sec:efficiency}). The axis itself is not new; our contribution lies in its systematic application to embedding models and in the distribution optimization built on it.
\item Distillation-free structural layer pruning: we prune the 22 layers of mmBERT-small to 4 / 13 layers and use the pruned models as base models, successfully training \model{bekko-embedding-v1-a8m} (7.67M AP) and \model{bekko-embedding-v1-a25m} (24.93M AP). Because no teacher model is required, this isolates how far pruning plus contrastive learning on public data alone can go (\secref{sec:pruning}). Our pruning-pattern ablation yields the observation that keeping the leading contiguous layers plus a deep Global layer (notably layer 18) rather than the final layer is effective (\secref{sec:layers}; we do not claim a causal localization of retrieval knowledge).
\item Multilingual, multi-domain data with two complementary LLM synthesizers: contrastive data covering 100+ languages and many domains, combining context-aware high-quality synthesis (Qwen3.5-35B-A3B) with fast synthesis by our own query-crafter-multilingual (\secref{sec:data}).
\item Single-GPU training and open release: smaller AP also shrinks training compute, and all training completes on a single GPU (\secref{sec:singlegpu}). The stage-1 data, hard negatives, and model weights are public.
\item Demonstrated on-device execution: quantizing the bulky vocabulary embedding table of the ONNX / OpenVINO artifacts to row-wise int8 (about 1/4 for the table alone) sharply shrinks the model file (a8m 124~MiB, about 1/3 of the fp32 distribution), with practical operation shown on Raspberry Pi 5, CPUs, and web browsers (\secref{sec:efficiency}).
\end{enumerate}

Paper organization: \secref{sec:related} related work; \secref{sec:design} design (three principles); \secref{sec:setup} evaluation setup; \secref{sec:results} retrieval quality; \secref{sec:analysis} analysis (pruning, recipe, QAT); \secref{sec:efficiency} efficiency and deployment; \secref{sec:limitations} limitations; \secref{sec:conclusion} conclusion. All numbers in this paper are based on frozen evaluation snapshots (\secref{sec:models}, \appref{app:details}).

\section{Related Work}
\label{sec:related}

\subsection{Bi-encoders and multilingual embeddings}
\label{sec:biencoder}

Sentence-BERT \citep{reimers2019sbert} recast BERT as a bi-encoder and established efficient sentence-embedding retrieval. For multilinguality, the two main lineages are knowledge distillation with parallel data \citep{reimers2020making} and large-scale multilingual pretraining exemplified by XLM-R \citep{conneau2020xlmr}. LaBSE \citep{feng2022labse} built a shared embedding space for 100+ languages, and Multilingual E5 \citep{wang2024me5} extended the English E5 recipe to multilingual small/base/large models. For Japanese, Ruri \citep{tsukagoshi2026ruri} showed that systematic pair design with synthetic data and reranker filtering yields high generality even at small scale. Bekko follows this line, pursuing competitiveness in the small-model band not through scale but through efficient architectural design and a data strategy modeled on Ruri.

\subsection{Data-centric multi-stage contrastive learning}
\label{sec:datacentric}

E5 \citep{wang2022e5} demonstrated the effectiveness of weakly supervised large-scale contrastive learning, and GTE \citep{li2023gte} systematized it as multi-stage training (large-scale weakly supervised pretraining plus multi-task supervised fine-tuning). GTE reports that contrastive pretraining performance saturates around a batch size of ten thousand, which grounds our batch-size choice (\secref{sec:recipe}). mGTE \citep{zhang2024mgte}, the multilingual extension of GTE, introduced RoPE and unpadding to enable 8192-token training. Qwen3 Embedding \citep{zhang2025qwen3embedding} reports the latest generation of LLM-based multi-stage training and adopts a masked contrastive loss that suppresses false negatives among in-batch negatives (the direct reference for our loss design, \secref{sec:recipe}). BGE-M3 \citep{chen2024bgem3} unifies dense/sparse/multi-vector retrieval in a single model and proposes token-length-grouped training with staged transitions (influencing our pipeline and token-length groups, \secref{sec:singlegpu}). FineWeb \citep{penedo2024fineweb} and the FineWeb-2 line \citep{messmer2025multilingual} showed that data-quality selection strongly shapes performance. LLM-based query synthesis for IR data augmentation was pioneered by InPars \citep{bonifacio2022inpars} and Promptagator \citep{dai2023promptagator}, and SWIM-IR \citep{thakur2024swimir} extended it to many languages (\secref{sec:data}).

\subsection{Efficient encoder design and model compression}
\label{sec:efficient}

ModernBERT \citep{warner2024modernbert} overhauled encoder-only design with modern optimizations---native 8192-token context, flash attention, unpadding---achieving Pareto improvements. mmBERT \citep{marone2025mmbert} pretrained that architecture on 3 trillion tokens of multilingual data. Bekko builds on this mmBERT-small.

The main approaches to model compression are knowledge distillation, exemplified by DistilBERT \citep{sanh2019distilbert}, and structural layer pruning (layer removal). Bekko uses the latter to construct its base models (one component of the design; \secref{sec:pruning}), with an established lineage of prior work. \citet{sajjad2023poormans} showed that layers can be dropped from pretrained encoders without distillation while largely preserving downstream performance. For decoder LLMs, ShortGPT \citep{men2024shortgpt} and \citet{gromov2024unreasonable} report that many deep layers are redundant and that light continued training after layer removal ``heals'' most of the degradation. We apply these findings to a multilingual retrieval encoder, and further examine which layers to keep on retrieval tasks (\secref{sec:layers}).

The closest prior work we are aware of is Granite Embedding Multilingual R2 \citep{granite2026multilingualr2}. Its small variant, granite-embedding-97m-multilingual-r2, builds on a ModernBERT-style multilingual encoder; relative to the large variant (311M, 22 layers), the paper describes downsizing by model pruning and vocabulary selection (12 layers; vocabulary reduced from the GPT-OSS-family 200K to 180K tokens), with weights initialized from the English 12-layer small model \citep{granite2026multilingualr2, granite2025r2}. Total parameters are 97M, and the paper's tables report 28M Active Parameters. For training, it combines contrastive learning with knowledge distillation from large decoder-based teachers---the same Mistral-7B-Instruct-family teacher as English R2 for English, and Granite-8B-family teachers for multilingual data \citep{granite2025r2, granite2026multilingualr2}. Bekko differs on three points: (a) no teacher model or distillation whatsoever; (b) more aggressive reduction (22 $\to$ 4 layers, just under 8M AP); and (c) smaller AP also reduces the compute needed for training, so all training completes on a single GPU in an ordinary workstation. EmbeddingGemma \citep{schechtervera2025embeddinggemma} reaches on-device SOTA at 106M AP from a Gemma 3 backbone via encoder-decoder initialization and geometric embedding distillation, but it also relies on distillation and its AP is an order of magnitude larger. Model2Vec / static embeddings \citep{tulkens2024model2vec, aarsen2025static} are CPU-fast but lack contextualization, placing them in a different quality band from Bekko.

Table~\ref{tab:positioning} positions Bekko among major multilingual embedding models. Bekko occupies a spot no other model in the table does: single-digit-million AP (7.67M), a contextual encoder, no distillation, and open release of both weights and multilingual training data. In particular, embedding models that release their large-scale multilingual pretraining (stage-1) data are, to our knowledge, rare, and this openness directly supports reproduction and extension of small multilingual model research.

\begin{table}[t]
\caption{Positioning among major multilingual embedding models (AP = non-embedding parameters). This table illustrates design positioning (distillation, openness) with representative examples; the unified model set for performance comparison (multilingual dense models with AP $\leq$ 350M + BM25, 13 models) is defined in \secref{sec:models}.}
\label{tab:positioning}
\centering
\footnotesize
\setlength{\tabcolsep}{2.4pt}
\begin{tabular}{lrrr>{\raggedright\arraybackslash}p{3.8cm}c>{\raggedright\arraybackslash}p{2.3cm}}
\toprule
Model & AP & Total & Max len & Use of distillation$^{\S}$ & Weights & Training data \\
\midrule
bekko-a8m & 7.67M & 106M & 8192 & none & \checkmark & \checkmark~(both stages$^{\diamond}$) \\
mE5-small & 21.6M & 118M & 512 & init only (MiniLM is distilled) & \checkmark & partial \\
bekko-a25m & 24.9M & 123M & 8192 & none & \checkmark & \checkmark~(both stages$^{\diamond}$) \\
granite-97m-r2 & 28.3M & 97M & 32768 & training (teacher-score distill.) & \checkmark & not released$^{\dagger}$ \\
embgemma-300m & 106.3M & 308M & 2048 & training (align to teacher emb.) & \checkmark & not released$^{\dagger\dagger}$ \\
gte-m-base & 113.3M & 305M & 8192 & --- & \checkmark & partial \\
bge-m3 & 311.8M & 568M & 8192 & self-distill.\ (unified score as teacher) & \checkmark & partial \\
Model2Vec$^{\ddagger}$ & --- (static) & tens of M & --- & distillation for staticization & \checkmark & ---$^{\ddagger}$ \\
\bottomrule
\end{tabular}
\begin{minipage}{\textwidth}
\footnotesize
\smallskip
$^{\S}$``Use of distillation'' distinguishes where teacher-derived distillation enters the construction process. mE5-small uses no distillation in its own training, but its initialization, multilingual MiniLM, is itself a distilled model. Granite R2 uses a distillation loss toward the similarity-score distributions of large teachers (Mistral-7B / Granite-8B-family models converted to embedders) during fine-tuning. EmbeddingGemma trains with distillation that directly aligns to the embedding space of the teacher Gemini Embedding. BGE-M3 uses not an external teacher but loss-level self-knowledge distillation in which the integrated dense/sparse/multi-vector score serves as the teacher signal (a separate mechanism from hard-negative mining). $^{\diamond}$Breakdown of Bekko's training-data release: stage 1 releases all of the data used, as a single public dataset; stage 2 releases the independently mined hard negatives, with the remainder consisting of already-public upstream datasets (Ruri v3 FT, BGE-family subsets, etc.; \appref{app:stage2})---i.e., every component of both stages is public data (stage 2 is not redistributed as a single dataset). $^{\dagger}$The training data is described as ``permissive, enterprise-friendly licensed'' but includes IBM-internal and IBM-generated synthetic data; the dataset itself is not released (model card, Data Collection). $^{\dagger\dagger}$Only checkpoints are released; the training data is outlined (web title--body pairs, a subset of Gecko's academic datasets, Gemini Embedding synthetic data, etc.) but the datasets and mixture ratios are not public. $^{\ddagger}$Model2Vec alone is a static embedding without context; all others are contextual encoders. Its staticization distillation uses only a vocabulary and a teacher model, with no contrastive data (the Tokenlearn pretraining used by the derived Potion models uses the public C4 corpus). All models in the table are multilingual (mE5 $\approx$100 languages; gte-m-base 70+; bge-m3 100+; granite has enhanced support for 52 languages, evaluated on 200+; embgemma and Bekko 100+; Model2Vec varies by variant).
\end{minipage}
\end{table}

\subsection{Choice of evaluation benchmarks}
\label{sec:benchmarks}

For evaluation, MTEB \citep{muennighoff2023mteb} established comprehensive evaluation and MMTEB \citep{enevoldsen2025mmteb} extended it to 500+ tasks and 250+ languages. We choose our evaluations to satisfy three requirements: (1) central claims must be verifiable on independent benchmarks that do not depend on datasets created by the author; (2) broad coverage of multilingual retrieval; and (3) lightweight enough to compare many models and settings under unified conditions. For lightweight retrieval evaluation there is NanoBEIR\footnote{\url{https://huggingface.co/collections/zeta-alpha-ai/nanobeir-66e1a0af21dfd93e620cd9f6}} (Zeta Alpha; \citealp{camara2024nanobeir}), which condenses each of the 13 BEIR \citep{thakur2021beir} datasets it covers to 50 queries with a small corpus, and machine-translated versions have been released by the community (LightOn, 8 languages \citep{sourty2025nanobeirmulti}; Liquid AI, Japanese/Korean \citep{liquidai2025nanobeir}; Serbian AI Society, Serbian \citep{serbianai2025nanobeirsr}; Sionic AI, Thai and Vietnamese \citep{sionic2025nanobeir})---neither NanoBEIR nor its translations were created by the author of this work.

We use official MMTEB Multilingual v2 and Multilingual NanoBEIR (MNanoBEIR; 14 languages; the original English NanoBEIR plus these third-party translations) as the primary evaluations, and HAKARI-Bench \citep{tateno2026hakari} for auxiliary comparison and as a development-time proxy. HAKARI-Bench is an IR benchmark created and released by the author of this work: it condenses multilingual, diverse tasks---including MMTEB retrieval tasks and BEIR---into small Nano subsets, comprising 500+ tasks, and allows many models to be compared cheaply under unified conditions. Its Nano rankings are reported to reproduce official rankings (on the common models and common tasks of MTEB, MMTEB, and BEIR) at Spearman $>$ 0.97.

\section{The Design of Bekko: Three Principles}
\label{sec:design}

\begin{figure}[t]
  \centering
  \includegraphics[width=\linewidth]{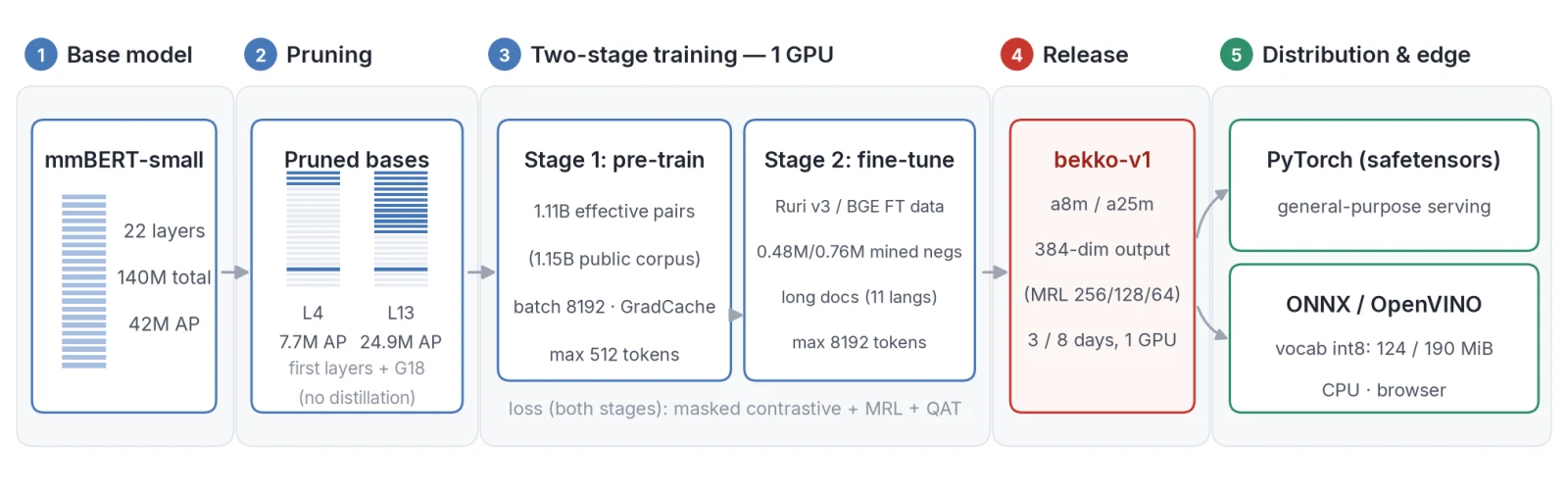}
  \caption{Bekko's construction pipeline. Without distillation, the pruned model serves as the base for two-stage contrastive training completed on a single GPU (a8m about 3 days / a25m about 8 days), and the ONNX / OpenVINO distributions achieve 124 / 190 MiB via int8 quantization of the vocabulary embedding (\secref{sec:pruning}--\secref{sec:singlegpu}, \secref{sec:int8}).}
  \label{fig:pipeline}
\end{figure}

Bekko's design concretizes the approach of \secref{sec:approach} into three principles. Principle 1: prune AP while preserving pretrained multilingual knowledge (structural pruning, \secref{sec:pruning}). Principle 2: compensate for the knowledge shortfall of a small model with broad data and two complementary synthetic datasets (\secref{sec:data}). Principle 3: extract quality with a modern contrastive-learning recipe (\secref{sec:recipe}). All of this runs on a single GPU (\secref{sec:singlegpu}). Figure~\ref{fig:pipeline} shows the overall picture.

\subsection{Base model: mmBERT-small}
\label{sec:mmbert}

We adopt jhu-clsp/mmBERT-small \citep{marone2025mmbert}, a multilingual encoder based on the ModernBERT architecture, with the following main specifications.

\begin{center}
\small
\begin{tabular}{ll}
\toprule
Item & Value \\
\midrule
Architecture & ModernBERT (encoder-only) \\
Total Parameters & 140M \\
Active Parameters (AP) & 42M \\
Layers & 22 \\
Hidden size & 384 \\
Attention heads & 6 \\
Intermediate size & 1152 \\
Vocabulary & 256{,}000 (Gemma 2 tokenizer) \\
Max position embeddings & 8{,}192 \\
Global attention interval & every 3rd layer \\
Local attention window & 128 \\
RoPE theta & 160{,}000 for both Global and Local \\
\bottomrule
\end{tabular}
\end{center}

\begin{figure}[t]
  \centering
  \includegraphics[width=\linewidth]{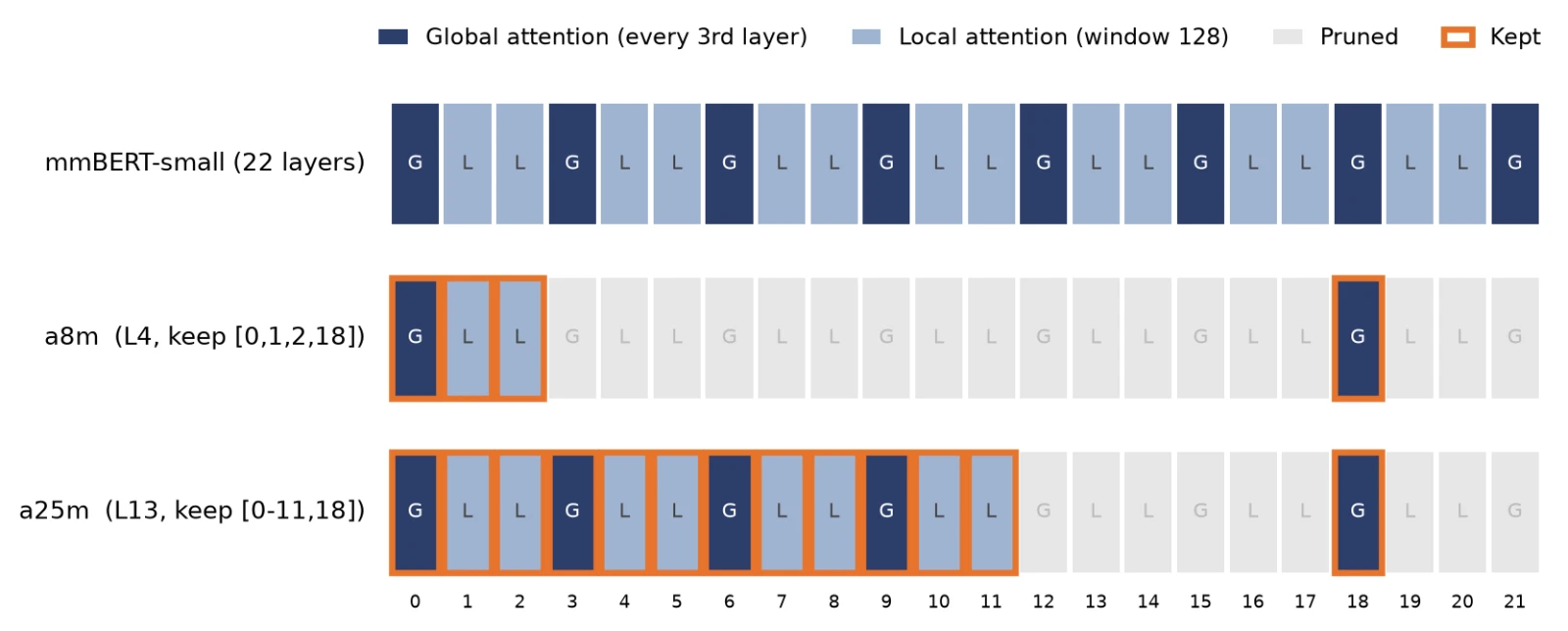}
  \caption{Structural layer pruning. We adopt designs that add the deep Global layer (18), not the final layer (21) (layer-selection comparison in \secref{sec:layers}).}
  \label{fig:pruning-layers}
\end{figure}

A key design property of ModernBERT is the repeating Global-Local-Local (G-L-L) pattern: of the 22 layers, layers 0, 3, 6, 9, 12, 15, 18, 21 use Global attention (G) over the full sequence, while the rest use Local attention (L) with a window of 128 (Figure~\ref{fig:pruning-layers}). Maintaining this rhythm becomes a design constraint for pruning (\secref{sec:pruning}). RoPE theta differs between ModernBERT and mmBERT: ModernBERT \citep{warner2024modernbert} uses 160{,}000 for Global layers and 10{,}000 for Local layers, whereas mmBERT sets both to 160{,}000, a more long-context-oriented configuration. This may contribute to long-input robustness from stage 1 onward (\secref{sec:longdoc}). We choose mmBERT-small for its broad language knowledge from pretraining on 3 trillion tokens across 1800+ languages, its small hidden size of 384, and ModernBERT's modern optimizations (flash attention, RoPE \citep{su2021roformer}).

\subsection{Principle 1: Structural layer pruning that preserves knowledge}
\label{sec:pruning}

From mmBERT-small's 22 layers we select a subset of layers useful for retrieval and construct pruned models by pure structural layer removal. The weights of retained layers are unchanged (no distillation or continued pretraining required).

Layer selection follows three principles: (1) keep the early contiguous layers, which carry low-level representations; (2) preserve the G-L-L rhythm as much as possible; and (3) include at least one deep Global layer for long-range dependencies. Because which pattern to keep strongly affects retrieval performance, we analyze it separately in \secref{sec:layers}. To preview the conclusion: configurations keeping a distant deep Global layer (18) rather than the final layer (21) perform well, and the advantage grows with the number of retained layers (\secref{sec:layers}). The released models' retained patterns and parameters are:

\begin{center}
\footnotesize
\setlength{\tabcolsep}{3pt}
\begin{tabular}{llllrr}
\toprule
Model & Base (pruned) & Retained layers & Layers & AP & Total \\
\midrule
\model{bekko-embedding-v1-a8m} & \model{mmBERT-L4H384-pruned} & [0, 1, 2, 18] & 4 & 7{,}671{,}168 & 105{,}975{,}168 \\
\model{bekko-embedding-v1-a25m} & \model{mmBERT-L13H384-pruned} & [0,1,2,\dots,11, 18] & 13 & 24{,}930{,}432 & 123{,}234{,}432 \\
\bottomrule
\end{tabular}
\end{center}

As noted in \secref{sec:ap-axis}, total parameters are dominated by the vocabulary embedding matrix ($\approx$98M), but inference cost is governed by AP (about 7.7M for a8m). The pruned models are not meant to be used as embedding models directly; they serve as the base models for contrastive training.

\subsection{Principle 2: Data to fill the knowledge gap}
\label{sec:data}

Bekko's data design generalizes the pair design of Ruri \citep{tsukagoshi2026ruri} to the multilingual setting. Ruri's data design has three key elements. (1) LLM-synthesized datasets---e.g., AutoWikiQA (about 2.5M items), generated by prompting an LLM to write questions and search queries from Japanese Wikipedia passages; Ruri feeds over 100M synthetic items into contrastive pretraining. (2) Large-scale unsupervised query--doc construction: mechanically extracting ``(section) title $\to$ paragraph/body'' pairs from Wikipedia, news, and academic sources, yielding asymmetric pairs without human annotation. (3) Large-scale unsupervised doc--doc construction: pair extraction treating different sentences or paragraphs of the same article as related text. Combining these into about 220M pairs for contrastive pretraining---with BM25 hard negatives, reranker filtering, and two-stage training---Ruri achieved high generality at small scale (for prior work on two-stage training and LLM synthesis themselves, see E5, GTE, InPars, Promptagator, and SWIM-IR in \secref{sec:datacentric}). We extend this (1)--(3) pair design to the multilingual setting.

\paragraph{Stage-1 data (released as \model{bekko-embedding-v1-unsupervised}).}
The stage-1 data totals about 1.15 billion pairs across 2{,}452 subsets, released on Hugging Face as bekko-embedding-v1-unsupervised\footnote{\url{https://huggingface.co/datasets/hotchpotch/bekko-embedding-v1-unsupervised}} (statistics are measured from the public data; \appref{app:stage1}). Training of the released models, after the bitext quality filter and block shaping described below, effectively used 1{,}863 subsets and about 1.11 billion rows (about 97\% of the public corpus). Pairs fall into two types: doc\_doc (symmetric document$\leftrightarrow$document pairs---translations, related paragraphs, etc.; 1{,}975 subsets, about 0.48B rows) and query\_doc (asymmetric pairs of a document and a short text pointing to it---search queries, article titles and headlines, long-form questions; 477 subsets, about 0.66B rows). Of these, 34 subsets are triplets with explicit hard negatives, but by role they belong to their respective pair types. The families are:

\begin{center}
\footnotesize
\setlength{\tabcolsep}{3pt}
\begin{tabular}{>{\raggedright\arraybackslash}p{2.0cm}>{\raggedright\arraybackslash}p{4.2cm}>{\raggedright\arraybackslash}p{1.5cm}r>{\raggedright\arraybackslash}p{4.3cm}}
\toprule
Family & Main sources & Type & Rows & Summary \\
\midrule
Parallel bitext & NLLB, CCMatrix / Opus-100 / WikiMatrix / WikiTitles & doc\_doc & 0.33B & 1{,}600+ language pairs (incl.\ non-English pairs). Part of the English bitext is a high-quality subset filtered by a reranker + embedding model \\
\addlinespace[3pt]
News & Multilingual CC-News (47 langs) / related-paragraph pairs (134 langs) & query\_doc / doc\_doc & 0.15B & headline$\to$body (query\_doc); paragraph pairs within/between articles (doc\_doc) \\
\addlinespace[3pt]
Wikipedia / FineWiki & finewiki (314 langs) / wikipedia-multilingual IR, lead, and paragraph pairs & query\_doc / doc\_doc & 0.19B & title$\to$body (query\_doc); related/lead-paragraph semantic-proximity pairs (doc\_doc) \\
\addlinespace[3pt]
Fast LLM synthesis & FineWeb2-IR (21 langs) / wikipedia-multilingual synthetic queries (11 langs) & query\_doc & 0.12B & synthesized by query-crafter-multilingual (below) \\
\addlinespace[3pt]
High-quality LLM synthesis & fineweb-edu / arXiv abstracts / PubMed abstracts / CC-News / English Wikipedia & query\_doc & 0.04B & synthesized by Qwen3.5-35B-A3B (below) \\
\addlinespace[3pt]
Existing IR pair collections & nomic unsupervised / lighton / swim-ir / all-nli / miracl, etc. & query\_doc / doc\_doc & 0.31B & existing weakly supervised pairs, QA, NLI \\
\bottomrule
\end{tabular}
\end{center}

The basis for calling Bekko a 100+-language model is the language coverage of this stage-1 data. The parallel bitext includes many-to-many translations from NLLB and CCMatrix across 1{,}622 normalized language pairs (about 0.33B rows); the Wikipedia family (finewiki) covers 314 languages and the news family 47--134 languages (\appref{app:stage1}). Cross-lingual alignment is also corroborated by the MMTEB BitextMining results (\secref{sec:mmteb}). Row counts are, however, skewed toward English and major languages, and low-resource-language performance remains limited (\secref{sec:limitations}).

Two preprocessing steps are applied at training time (details in \appref{app:stage1}). (1) Bitext quality filter: the many-to-many bitext from NLLB and CCMatrix contains low-quality machine-mined translation pairs, so we sampled and audited translations per language pair (subset), scored their quality, and excluded blatantly low-quality language pairs---leaving about 1{,}300 of the 1{,}575 many-to-many language-pair subsets in use (the 1{,}622 normalized pairs above count unique pairs across all parallel-bitext families). Separately, the high-quality English bitext subset is built by row-level filtering with a reranker and an embedding model (\appref{app:stage1}). (2) Deduplicated blocks: since duplicate texts within a batch become false in-batch negatives, within each subset we pre-build randomly sampled 8{,}192-row blocks with no query/doc duplication inside a block, and each block is used as one batch. Moreover, in both stages, every batch is always drawn from a single subset---subsets are never mixed within a batch; the subset for each batch is chosen with probability proportional to row count. This keeps in-batch negatives from a single source and distribution, stabilizing negative difficulty.

Two complementary synthetic datasets. LLM query synthesis from documents was pioneered by InPars \citep{bonifacio2022inpars} and Promptagator \citep{dai2023promptagator} and extended multilingually by SWIM-IR \citep{thakur2024swimir}; Ruri's AutoWikiQA is in the same family. Bekko does not claim novelty in synthetic query generation itself; its distinguishing point is combining two complementary synthesizers---quality-oriented and speed-oriented---at multilingual scale (about 0.16B pairs in total):

\begin{enumerate}[leftmargin=*]
\item High-quality synthesis (Qwen3.5-35B-A3B): for documents from fineweb-edu, arXiv abstracts, PubMed abstracts, CC-News, and English Wikipedia, generate IR queries with Qwen3.5-35B-A3B. Query form, perspective, and specificity (direct / partial / with typos / keywords-only / ambiguous, etc.) and the assumed asker are controlled per row, producing diverse, realistic queries that engage with the context. About 39M pairs in total (English Wikipedia $\approx$29.4M, arXiv $\approx$2.88M, PubMed $\approx$2.36M, etc.; generation details in \appref{app:stage1}).
\item Fast synthesis (query-crafter-multilingual\footnote{\url{https://huggingface.co/hotchpotch/query-crafter-multilingual}}): our own multilingual query-generation model (Qwen3-1.7B-based) mass-produces queries for FineWeb2-IR (21 languages) and others. It generates seven query types---keywords, natural-language queries, titles, FAQs, summaries, etc.---with weighted mixing. It is light and scales to volume, but is weaker at complex queries requiring deep reading (\appref{app:querycrafter}). About 0.12B pairs.
\end{enumerate}

\paragraph{Stage-2 data (hard negatives released as \model{bekko-embedding-v1-hard-negatives}).}
The stage-2 data consists of high-quality rows in query / positive (1) / hard-negatives format; our own mined portion is released on Hugging Face as bekko-embedding-v1-hard-negatives\footnote{\label{fn:hn}\url{https://huggingface.co/datasets/hotchpotch/bekko-embedding-v1-hard-negatives}}.

\begin{center}
\small
\setlength{\tabcolsep}{4pt}
\begin{tabular}{>{\raggedright\arraybackslash}p{2.5cm}>{\raggedright\arraybackslash}p{6.2cm}>{\raggedright\arraybackslash}p{2.0cm}>{\raggedright\arraybackslash}p{2.8cm}}
\toprule
Family & Data & Type & Summary \\
\midrule
Japanese & Ruri v3 FT \citep{tsukagoshi2026ruri} & query\_doc & all Ruri FT subsets (max 512 tokens) \\
\addlinespace[3pt]
BGE subsets & dureader, hotpotqa, miracl, mr-tydi, squad, msmarco ($\leq$200k rows), Chinese mmarco ($\leq$80k rows), pubmed\_qa\_labeled (query\_doc) / ATEC, BQ, LCQMC, PAWSX (doc\_doc) & query\_doc / doc\_doc & BGE-M3 FT subsets with relatively clear licenses \\
\addlinespace[3pt]
Own hard negatives & bekko-embedding-v1-hard-negatives (759{,}587 rows, 22 subsets) & query\_doc / doc\_doc & see below \\
\bottomrule
\end{tabular}
\end{center}

Our mined hard negatives comprise general English IR (wikipedia\_hard\_negatives\_english 500k rows; agnews, gooaq, natural\_questions, trivia\_qa), doc\_doc (all\_nli, coco\_captions), domain data (arxiv, codesearch, fineweb, pubmed), and long-document versions (\model{wikipedia\_hard\_negatives\_long\_docs\_*}, 11 languages). Each row has a query, one positive, and up to 15 hard negatives (up to 7 used in FT). Mining uses hybrid sparse BM25 + dense retrieval for short-to-medium texts, where the dense model is an earlier development checkpoint of our own model (256-dimension truncated)---a self-bootstrapping setup. Quality filters such as positive exclusion and near-duplicate removal are applied. The long-document versions (max 8192 tokens) are mined with sparse BM25 only, with queries generated by an LLM (DeepSeek-V4, \model{deepseek-v4-flash}), following the MLDR framework \citep{chen2024bgem3}. Documents are split-sampled so that only part of a long document contains the answer, forming hard negatives that endow stage 2 with 8192-token long-context retrieval ability.

Family-level breakdowns, examples, and tendencies of the stage-1 data (parallel bitext, broad-positive doc\_doc, the two synthetic query\_doc lines) are detailed in \appref{app:stage1}. The stage-2 mixture is identical for a8m and a25m (37 subsets, 1{,}780{,}001 effective rows, including MS MARCO $\leq$200k rows and Chinese mMARCO $\leq$80k rows from the BGE-M3 FT data), and 482{,}000 rows are sampled with per-subset caps from the 759{,}587 released hard-negative rows (full composition and counts in \appref{app:stage2}).

\subsection{Principle 3: Training recipe (two stages, loss, MRL)}
\label{sec:recipe}

We train with two-stage contrastive learning. Stage 1 builds general-purpose representations on multilingual weakly supervised pairs (max\_seq 512, batch 8192), following the insight that large-scale contrastive learning on weakly supervised pairs produces high-quality embeddings \citep{wang2022e5, neelakantan2022text}. Stage 2 follows the multi-stage training of GTE \citep{li2023gte}, BGE-M3 \citep{chen2024bgem3}, and Ruri \citep{tsukagoshi2026ruri}: it raises retrieval quality with high-quality data including hard negatives and adapts to long context (max\_seq 8192). The final hyperparameters for a8m / a25m are:

\begin{center}
\small
\setlength{\tabcolsep}{3pt}
\resizebox{\textwidth}{!}{%
\begin{tabular}{lllll}
\toprule
Parameter & Stage 1 a8m & Stage 1 a25m & Stage 2 a8m & Stage 2 a25m \\
\midrule
Base model & \model{mmBERT-L4H384-pruned} & \model{mmBERT-L13H384-pruned} & stage-1 model & stage-1 model (ckpt merge; see text) \\
Max tokens & 512 & 512 & 512 / 8192 (per data group) & 512 / 8192 \\
Batch size & 8192 & 8192 & 1152 / 192 (per data group) & 1152 / 192 \\
Learning rate & $9.0\times10^{-4}$ & $3.0\times10^{-4}$ & $1.0\times10^{-5}$ & $4.0\times10^{-6}$ \\
Warmup ratio & 0.1 & 0.1 & 0.3 & 0.3 \\
Epochs & 1 & 1 & 1 & 1 \\
Temperature $\tau$ / margin $m$ & 0.03 / 0.1 & 0.03 / 0.1 & 0.03 / 0.01 & 0.03 / 0.01 \\
MRL dims / weights & \multicolumn{4}{l}{[384,256,128,64] / [1.0,0.3,0.15,0.1] (same in all runs)} \\
Gradient-cache mini-batch & 2048 & 512 & 16--128 & 16--128 \\
\bottomrule
\end{tabular}}
\end{center}

The learning rate is set higher for smaller models (a8m stage 1 $9.0\times10^{-4}$ $>$ a25m $3.0\times10^{-4}$), a setting based on development-time exploration.

Stage-2 initialization: a8m starts from the final stage-1 checkpoint as is. a25m starts from a linear merge (weights 0.5/0.5, bf16) of the final stage-1 checkpoint and the intermediate checkpoint with the best development evaluation during training---an average of the weights of two checkpoints from the same run, adopted as the best initialization by comparison on the development evaluation.

Prefix: the released models attach no prefix to either queries or documents. This departs from the role prefixes of E5 \citep{wang2022e5} and Ruri \citep{tsukagoshi2026ruri}, prioritizing the simplicity of handling symmetric and asymmetric tasks through the same API. In a controlled comparison on a8m stage 1, the prefix conditions differed by at most about 0.003 on the development evaluation, with no clear advantage for role prefixes (\secref{sec:recipeablation}).

Masked contrastive loss. As the base loss we use the masked contrastive loss proposed by Qwen3 Embedding \citep{zhang2025qwen3embedding}; the per-pair-type choice of loss direction and the treatment of explicit hard negatives described below are our adaptation. In in-batch contrastive learning over query--document pairs $\{(q_i, d_i)\}_{i=1}^{N}$ ($N$ the batch size), the negatives for $q_i$ are the other documents $\{d_j\}_{j\neq i}$ (in-batch negatives). At large batch sizes, however, these ``negatives'' increasingly contain documents that are actually valid answers for $q_i$ (false negatives). The masked loss suppresses them with the rule: a negative candidate whose score exceeds the positive score by more than a margin $m$ is likely a false negative and is removed from the denominator. With $\mathrm{sim}$ cosine similarity and $\tau$ the temperature:

$$
L_{q \rightarrow d} = -\frac{1}{N}\sum_{i=1}^{N}\log \frac{\exp(\mathrm{sim}(q_i, d_i)/\tau)}{\exp(\mathrm{sim}(q_i, d_i)/\tau) + \sum_{j \neq i} M_{ij}\,\exp(\mathrm{sim}(q_i, d_j)/\tau)}
$$

The mask $M_{ij}\in\{0,1\}$ is given below; masked-out terms ($M_{ij}=0$) are implemented by setting the logit to $-\infty$ ($\exp(-\infty)=0$), removing them from the denominator:

$$
M_{ij} = \mathbf{1}\left[\mathrm{sim}(q_i, d_j) \leq \mathrm{sim}(q_i, d_i) + m\right]
$$

The composition of the partition function (denominator) follows Qwen3 Embedding's expanded normalization factor, which stands in the improved-contrastive-loss lineage of GTE \citep{li2023gte}: besides the other documents $d_j$, it includes the query--query similarities $q_i$--$q_j$ and document--document similarities $d_i$--$d_j$, with the same margin rule applied to every term (the equation shows only the q--d term). Our treatment of explicit hard negatives, however, differs from that of Qwen3 Embedding, which applies the mask to hard negatives as well: the explicit hard negatives assigned to a query (the negative list of a triplet) are verified negatives, so we exempt them from the masking rule in that query's loss and always keep them in the denominator (negatives assigned to other queries may still be masked when they appear as in-batch candidates). The direction of the loss branches by pair type: symmetric doc\_doc pairs (translations, NLI, Wikipedia, etc.) use the bidirectional loss $L_{\mathrm{bidir}} = (L_{q\rightarrow d} + L_{d\rightarrow q})/2$, which averages the loss with the roles of $q$ and $d$ swapped, while asymmetric query\_doc pairs (synthetic IR, retrieval) use the unidirectional loss. Temperature is $\tau=0.03$; the margin is $m=0.1$ in stage 1 and $m=0.01$ in stage 2. Since false-negative contamination matters most in stage 2, which contains hard negatives, we expect the mask's role to be larger there (we did not collect statistics of mask activation per stage).

Large batches and GradCache. Large-batch contrastive learning raises the chance that in-batch negatives include hard negatives, substantially boosting retrieval quality \citep{neelakantan2022text}. To achieve this on a single GPU we use GradCache \citep{gao2021gradcache}, which decouples embedding computation from backpropagation and splits the batch into small mini-batches (2048 for a8m stage 1) while caching gradients, computing full-batch in-batch negatives without approximation; a practical implementation is provided as the Cached losses of the Sentence Transformers library \citep{reimers2019sbert} (the \model{CachedMultipleNegativesRankingLoss} family\footnote{\url{https://sbert.net/docs/package_reference/sentence_transformer/losses.html}}). Our masked loss is implemented with the same gradient-caching structure, handling batch size 8192 on one GPU (each batch is drawn from a single subset; \secref{sec:data}). We adopt 8192 because GTE \citep{li2023gte} reports that contrastive-pretraining performance saturates around a batch size of ten thousand.

MRL. MatryoshkaLoss \citep{kusupati2022matryoshka} applies the contrastive loss simultaneously at multiple dimensionalities, learning multi-granularity embeddings whose dimensions can be truncated at deployment (dims [384,256,128,64], weights [1.0,0.3,0.15,0.1]). Lower dimensions get smaller weights to protect full-dimension quality. MRL is especially useful for ultra-compact models, since low-resource deployments often want smaller output vectors as well (the released models lose only 1.3--1.8\% overall when truncated to 256 dimensions; \secref{sec:vectorcompress}). In a weight comparison at the release dims (a proxy ablation on an a8m-like configuration with 20\% of the data; \appref{app:ablations}), the spread across settings was as small as about 0.001 on the HAKARI overall average, and the two settings with somewhat stronger low-dimension weights slightly beat the released choice (the released value was last of the three, by a slim margin)---weight choice is a low-sensitivity knob.

The total loss adds an auxiliary quantization-aware training (QAT) term. QAT here means computing the same contrastive loss on embeddings pseudo-quantized to int8 / ubinary during training, teaching the representation to survive quantization in advance. Unlike common QAT, which quantizes weights and activations, ours targets the output embedding space (model weights are never quantized during training; post-training weight quantization is treated separately in \secref{sec:weightquant}; implementation and observations in \secref{sec:qat}). The implementation sums, inside each MRL dimension, the float loss and the quantized-embedding losses with weights:

$$
L = \sum_{k \in \{384,256,128,64\}} w^{\mathrm{MRL}}_k \left( L^{\mathrm{mask}}_{[:k]} \;+\; \sum_{p \in \{\mathrm{int8},\,\mathrm{ubinary}\}} w^{\mathrm{QAT}}_p\, L^{\mathrm{mask}}_{p,[:k]} \right)
$$

where $L^{\mathrm{mask}}_{[:k]}$ is the loss on the first $k$ dimensions and $L^{\mathrm{mask}}_{p,[:k]}$ the same loss on embeddings pseudo-quantized by $p$ at those $k$ dimensions, with $w^{\mathrm{MRL}}=[1.0,0.3,0.15,0.1]$ and $w^{\mathrm{QAT}}=[0.1,0.1]$ (the effective weight of a QAT term is the product $w^{\mathrm{MRL}}_k \cdot w^{\mathrm{QAT}}_p$). The QAT terms are auxiliary; observations and interpretation are given in \secref{sec:qat}.

\subsection{Single-GPU training and token-length groups}
\label{sec:singlegpu}

All training ran on a single GPU (NVIDIA RTX PRO 6000 Blackwell Max-Q, 96GB GPU memory): about 3 days for a8m and about 8 days for a25m. This is enabled by the small AP itself, plus GradCache's memory-efficient large batches (\secref{sec:recipe}), padding reduction via token-length-grouped batch settings, and ModernBERT's flash attention. Stage 1 trains at max 512 tokens with a fixed batch of 8192. Stage 2, following BGE-M3's token-length-grouped training \citep{chen2024bgem3}, splits data into two token-length groups with separate batch sizes: short-to-medium data (max 512 tokens) uses batch 1152, and long-document data (max 8192 tokens) uses batch 192. Each stage-2 training example consists of one query, one positive, and up to 7 hard negatives. Large embedding models are often trained on distributed clusters of tens of GPUs---BGE-M3 up to 96 GPUs (A800), mGTE 32 GPUs (A100), E5 up to 64 GPUs (V100) \citep{chen2024bgem3, zhang2024mgte, wang2022e5}---whereas Bekko completes on one workstation GPU. On a consumer single GPU (RTX 5090), shrinking the gradient-cache mini-batch should reduce the required GPU memory and allow the same training at somewhat longer wall-clock time, though we have not run the full training on an RTX 5090.

\section{Evaluation Setup}
\label{sec:setup}

\subsection{Evaluation benchmarks}
\label{sec:evalbench}

Primary evaluations---two tracks that do not depend on author-created datasets. (1) MMTEB Multilingual v2 \citep{enevoldsen2025mmteb}: 131 tasks; the primary source for the official evaluation of a8m/a25m (\secref{sec:mmteb}), with Retrieval \nDCG{} as the headline metric. (2) Multilingual NanoBEIR (14 languages): the original English NanoBEIR (by Zeta Alpha; itself a condensation of BEIR \citep{thakur2021beir}) plus its community translations (\secref{sec:benchmarks}; none of the datasets are by this author). Measurement runs under the unified conditions of the HAKARI-Bench harness (\secref{sec:mnanobeir}). Note that the NanoBEIR family is a lightweight approximation with about 50 queries per task and does not replace full-scale evaluation---hence official MMTEB always accompanies it as primary evaluation (1).

Auxiliary comparison and proxy---HAKARI-Bench \citep{tateno2026hakari}: a lightweight multilingual IR benchmark that Nano-izes MTEB/MMTEB/BEIR. It measures NanoMMTEB-v2, NanoRTEB, NanoMLDR (long documents), NanoLongEmbed (long inputs), NanoCoIR (code), and per-language MTEB variants (e.g., Japanese NanoJMTEB) under unified conditions. We use it for auxiliary comparison of the released models (\secref{sec:mnanobeir}, efficiency-frontier visualization) and for development-time proxy evaluation and design ablations (\secref{sec:analysis}). HAKARI's dense evaluation tries both cosine and dot scoring per task and adopts the higher one (a per-task best-of-similarity upper bound, stated in the HAKARI paper)---the uplift for Bekko itself from this choice is below +0.0002 versus fixed cosine, but all HAKARI-derived numbers and rankings in this paper are values under that protocol. Because HAKARI-Bench is a benchmark released by this author (self-citation), central retrieval-quality claims are always cross-checked on the independent official MMTEB Multilingual v2 (\secref{sec:mmteb}).

Inference-efficiency and quantization evaluation---the quality retention of model-weight quantization (vocabulary-only int8 / full int8; \secref{sec:weightquant}) is evaluated on NanoBEIR-en + NanoMMTEB-v2, with input-length-dependent effects checked on NanoLongEmbed, NanoMLDR, and NanoCoIR; inference speed on CPU / Raspberry Pi / Apple Silicon / CUDA GPU is measured by encoding Natural Questions documents (\secref{sec:speed}), and the browser (WebGPU / WASM) is measured on the same Natural Questions document encoding (\secref{sec:browser}).

\subsection{Compared models}
\label{sec:models}

As stated in \secref{sec:ap-axis}, inference compute is governed by AP (non-embedding parameters). We unify the comparison set by the criterion ``multilingual dense embedding models with openly released weights under commercially usable licenses and AP $\leq$ 350M,'' plus the lexical baseline BM25. This criterion excludes non-commercial-license models (e.g., jina-embeddings-v3), decoder-based LLM-adapted embedders whose AP exceeds the bound (e.g., harrier-oss-v1-0.6b, AP 440M), and static embeddings without contextualization (\secref{sec:efficient}). harrier-oss-v1-270m \citep{microsoft2026harrier} is an embedding model with a decoder-only architecture (last-token pooling), but it meets all criteria---commercial use, released weights, multilingual dense, AP $\leq$ 350M---and is therefore included. The AP-300M-class models (mE5-large, bge-m3, arctic-embed-l-v2.0 \citep{yu2024arctic}) serve as the upper-side references for how far Bekko, with 1--2 orders of magnitude less AP, can close the gap. BM25 has no MMTEB score and is used only in the HAKARI-side comparisons (\secref{sec:mnanobeir}--\secref{sec:longdoc}). The BM25 scores are computed by evaluating the BM25 candidate rankings saved when each Nano set was built---the candidate-construction pipeline is based on Okapi BM25 of the bm25s library, with language-specific tokenization (word segmenters for Japanese, Chinese, Thai, Korean, and Vietnamese; stemmers where available; regex otherwise). Below, the multilingual-e5 family is abbreviated as in mE5-small, and gte-multilingual-base as gte-m-base.

\begin{center}
\small
\begin{tabular}{lrrcr}
\toprule
Model & AP & Total & Dims & Max input \\
\midrule
BM25 (lexical; \secref{sec:mnanobeir}--\secref{sec:longdoc} only) & --- & --- & --- & --- \\
bekko-embedding-v1-a8m & 7.67M & 106M & 384 & 8192 \\
mE5-small & 21.6M & 118M & 384 & 512 \\
bekko-embedding-v1-a25m & 24.93M & 123M & 384 & 8192 \\
granite-97m-r2 & 28.3M & 97M & 384 & 32768 \\
mE5-base & 86.0M & 278M & 768 & 512 \\
harrier-oss-v1-270m & 100.3M & 268M & 640 & 32768 \\
embeddinggemma-300m & 106.3M & 308M & 768 & 2048 \\
granite-311m-r2 & 110.3M & 312M & 768 & 32768 \\
gte-m-base & 113.3M & 305M & 768 & 8192 \\
mE5-large & 303.9M & 560M & 1024 & 512 \\
arctic-embed-l-v2.0 & 311.8M & 568M & 1024 & 8192 \\
bge-m3 & 311.8M & 568M & 1024 & 8192 \\
\bottomrule
\end{tabular}
\end{center}

AP is computed from MMTEB's \model{n\_active\_parameters\_override} when present, otherwise \model{n\_parameters} $-$ \model{n\_embedding\_parameters}. That mE5-small is 118M in total but only 21.6M in AP epitomizes why this axis matters.

Fairness of comparison: in official MMTEB (\secref{sec:mmteb}), all competitor models completed all 131 tasks in the same snapshot (2026-06-28), and each model is evaluated with its own recommended prompt and dtype as registered in MTEB's \model{ModelMeta}. Competitor scores come from the official MMTEB cache; for Bekko, we evaluated the released models (no prefix; \secref{sec:recipe}) on the identical task set (131 tasks) with the identical aggregation rules. To avoid depending on AP alone, we also report total parameters, distribution size, and throughput (\secref{sec:speed}).

\section{Results: Retrieval Quality with Tiny Active Parameters}
\label{sec:results}

We take official MMTEB Multilingual v2 (\secref{sec:mmteb}) and Multilingual NanoBEIR over third-party datasets (14 languages, \secref{sec:mnanobeir}) as primary evaluations, and use the author-released lightweight HAKARI-Bench (\secref{sec:mnanobeir}) as auxiliary comparison for efficiency-frontier visualization and multi-angle checks. Note that although a8m and a25m share identical stage-2 data (\secref{sec:data}), they differ not only in depth but also in learning rate, gradient-cache settings, and a25m's checkpoint merge (\secref{sec:recipe}); comparisons between a8m and a25m in this paper are therefore comparisons of the two released configurations, not a controlled scaling comparison that varies AP (capacity) alone.

\subsection{Official MMTEB Multilingual v2 (primary evaluation)}
\label{sec:mmteb}

We evaluated a8m and a25m on official MMTEB Multilingual v2 (131 tasks). Competitor values come from the official 2026-06-28 snapshot (mteb 2.16.1; the 170 models that completed all 131 tasks); Bekko values come from evaluating the released models on the identical task set with identical aggregation rules (task score = mean over splits/subsets; per-type = mean over tasks of that type) (\secref{sec:models}). Table~\ref{tab:mmteb} reports per-task-type scores for the unified comparison set. Since retrieval is this paper's focus, Retrieval (Ret) is the headline column, alongside the overall task mean (Mean) and all other task types. For reference, the rows a8m-pt / a25m-pt give the pretrained models at the end of stage 1 (large-scale contrastive learning).

\begin{table}[t]
\caption{Official MMTEB Multilingual v2 per-task-type scores (all $\times$100; ascending AP). Mean = mean of 131 tasks (Mean(Task)); Ret = Retrieval (task-macro, \nDCG{}); Rerank = Reranking; Bitext = BitextMining; STS = Semantic Textual Similarity; PairCls = PairClassification; Class = Classification; Clust = Clustering; MultiLbl = MultilabelClassification. InstructionReranking is omitted (centered scores near 0 for all models). The -pt rows are stage-1-only reference values. Bold marks scores where Bekko is strong, not column bests (see the highlights and body text).}
\label{tab:mmteb}
\centering
\small
\setlength{\tabcolsep}{2.2pt}
\begin{tabular}{lrrrrrrrrrrr}
\toprule
Model & AP & dims & Mean & Ret & Rerank & Bitext & STS & PairCls & Class & Clust & MultiLbl \\
\midrule
bekko-a8m-pt & 7.7M & 384 & 55.0 & 55.1 & 42.3 & 72.1 & 68.6 & 75.6 & 54.4 & 41.4 & 18.1 \\
bekko-a8m & 7.7M & 384 & \textbf{56.7} & \textbf{56.2} & 60.6 & \textbf{73.1} & 71.6 & 76.1 & 54.8 & 42.4 & 18.6 \\
mE5-small & 21.6M & 384 & 56.4 & 50.9 & 60.4 & 69.4 & 71.7 & 77.3 & 57.0 & 41.5 & 19.3 \\
bekko-a25m-pt & 24.9M & 384 & 56.7 & 56.1 & 44.4 & 74.1 & 70.6 & 76.7 & 56.9 & 42.5 & 18.6 \\
bekko-a25m & 24.9M & 384 & \textbf{58.3} & \textbf{57.5} & \textbf{61.6} & \textbf{75.4} & \textbf{73.4} & 77.4 & 57.2 & 43.0 & 19.2 \\
granite-97m-r2 & 28.3M & 384 & 51.9 & 60.3 & 59.4 & 44.2 & 65.6 & 74.5 & 49.6 & 43.6 & 17.9 \\
mE5-base & 86.0M & 768 & 57.0 & 52.7 & 60.2 & 69.5 & 71.4 & 77.2 & 58.2 & 41.8 & 20.2 \\
harrier-270m & 100.3M & 640 & 66.6 & 66.4 & 61.9 & 81.5 & 75.4 & 80.1 & 70.8 & 52.5 & 24.0 \\
embgemma-300m & 106.3M & 768 & 61.2 & 62.5 & 63.3 & 64.4 & 74.7 & 81.4 & 60.9 & 51.2 & 24.8 \\
granite-311m-r2 & 110.3M & 768 & 56.0 & 65.2 & 62.0 & 57.9 & 69.0 & 76.1 & 53.3 & 44.4 & 18.5 \\
gte-m-base & 113.3M & 768 & 58.3 & 57.2 & 60.7 & 71.8 & 72.9 & 80.5 & 57.2 & 44.2 & 19.8 \\
mE5-large & 303.9M & 1024 & 58.6 & 53.7 & 62.9 & 73.8 & 73.3 & 78.8 & 59.4 & 42.7 & 21.1 \\
arctic-l-v2.0 & 311.8M & 1024 & 57.0 & 58.4 & 63.7 & 64.1 & 70.1 & 76.8 & 57.4 & 42.8 & 18.9 \\
bge-m3 & 311.8M & 1024 & 59.6 & 54.6 & 62.8 & 79.1 & 74.1 & 80.8 & 60.3 & 40.9 & 20.1 \\
\bottomrule
\end{tabular}
\end{table}

The highlights are as follows.

\begin{itemize}[leftmargin=*]
\item Retrieval (task-macro): Bekko matches or surpasses models 1--2 orders of magnitude larger in AP. a8m at 56.2 (7.67M AP) beats mE5-small (50.9), mE5-base (52.7), mE5-large (53.7), and bge-m3 (54.6)---models with 3--40$\times$ its AP. a25m at 57.5 further reaches parity with gte-m-base (57.2) and comes within one point of arctic-embed-l-v2.0 (58.4, 12$\times$ AP). Above Bekko in the unified set are only the strongly retrieval-specialized granite R2 family (97m 60.3 / 311m 65.2), embeddinggemma-300m (62.5), harrier-270m (66.4), and arctic. Bekko's strength concentrates in Retrieval.
\item Overall (Mean): a25m (58.3) ties gte-m-base (58.3), surpasses arctic-l-v2.0 (57.0, 12$\times$ AP), mE5-base (57.0), and the granite family, and comes within 0.3 of mE5-large (58.6), but does not reach bge-m3 (59.6), embeddinggemma (61.2), or harrier-270m (66.6). a8m (56.7) beats mE5-small (56.4). The relation ``a25m matches or surpasses bge-m3'' thus holds on MMTEB Retrieval, MNanoBEIR, and HAKARI Overall (within 0.01 on the latter two), but not on the MMTEB overall mean.
\item Per task type (Table~\ref{tab:mmteb}): strong on retrieval-like types (Ret, Rerank), cross-lingual alignment (Bitext), and STS; mid-pack on PairCls, Clust, and MultiLbl; somewhat low on Class.
\item Cross-lingual alignment (Bitext): on BitextMining (task-macro), a25m scores 75.4, third in the unified set after harrier-270m (81.5) and bge-m3 (79.1), above mE5-large (73.8), gte-m-base (71.8), arctic (64.1), and the granite R2 family (311m 57.9 / 97m 44.2). a8m at 73.1 sits within 0.7 of mE5-large, which has 40$\times$ its AP. This aligns with the parallel bitext in the stage-1 data (1,600+ language pairs, about 0.33B rows; \secref{sec:data}). The primary basis for the 100+-language claim remains the data's language coverage (\secref{sec:data}); BitextMining corroborates cross-lingual consistency within the language pairs the benchmark covers.
\end{itemize}

The central claim of this paper concerns quality as first-stage dense retrieval; we do not claim state of the art in general-purpose sentence embedding including Clustering and other types.

\begin{figure}[t]
  \centering
  \includegraphics[width=\linewidth]{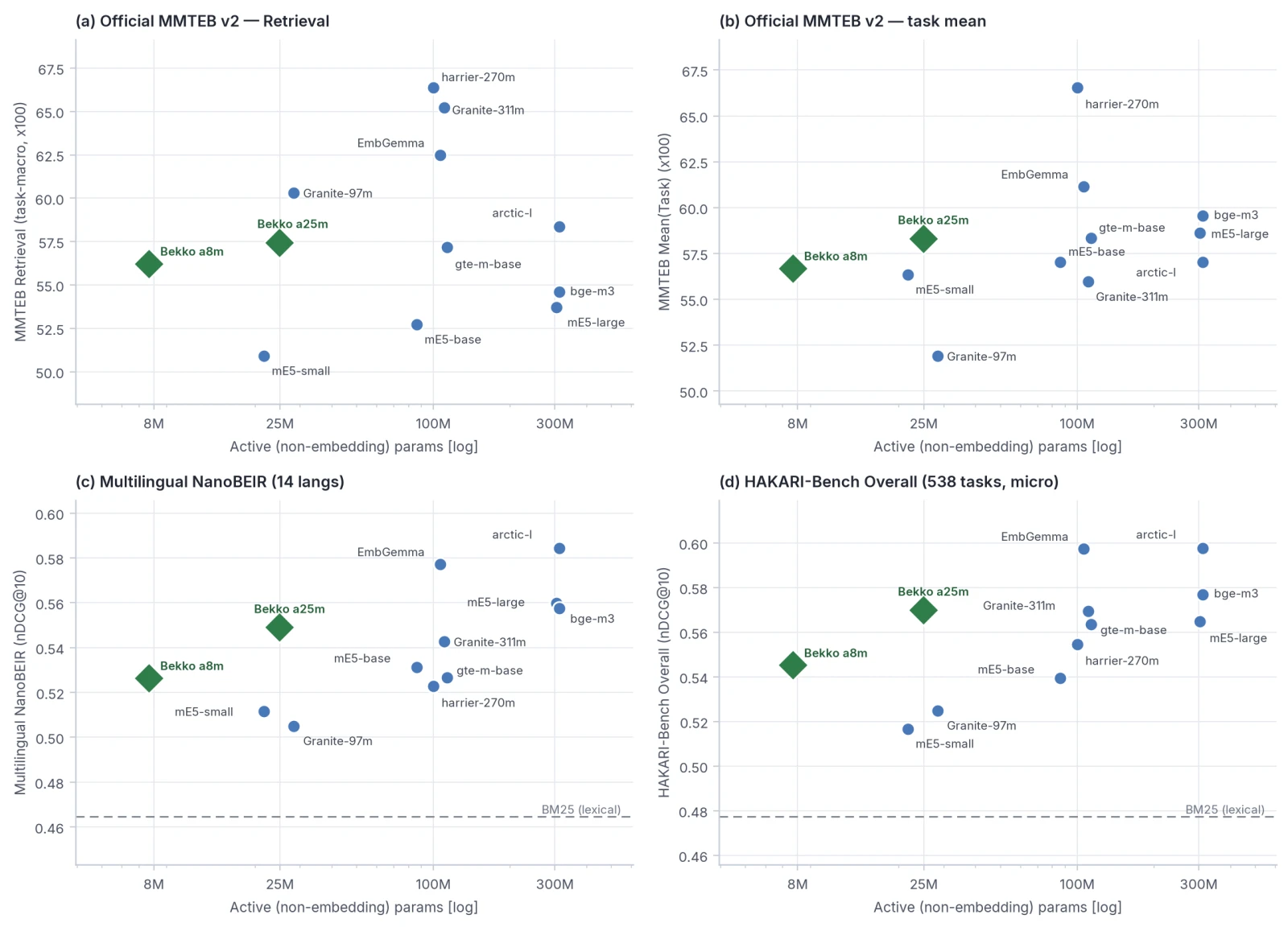}
  \caption{Retrieval quality versus AP (log scale; the 12 dense models of the unified set; 2$\times$2 panels)---the central evidence for the AP axis (\secref{sec:ap-axis}). (a) On MMTEB Retrieval, a8m (7.7M AP) beats bge-m3 and the mE5 family, and a25m approaches arctic (12$\times$ AP). (b) On the MMTEB overall mean, a25m also exceeds arctic and ties gte-m-base (Table~\ref{tab:mmteb}). (c) On Multilingual NanoBEIR (\secref{sec:mnanobeir}), a25m lands within 0.02 of the 300M-class bge-m3 and mE5-large, and every dense model beats BM25 (dashed). (d) HAKARI-Bench Overall shows broadly the same picture.}
  \label{fig:pareto}
\end{figure}

Per-language trends are checked with the per-language scores of primary evaluation (2), Multilingual NanoBEIR (\secref{sec:mnanobeir}, \appref{app:details}, Figure~\ref{fig:perlang}). a25m beats mE5-small and BM25 in all 14 languages and even bge-m3 in English, Portuguese, and Arabic, while trailing bge-m3 and mE5-large by small margins elsewhere (at most 0.026 vs.\ bge-m3). In other words, it runs alongside the 300M class (about 12$\times$ AP) across languages with small per-language dips, leaving headroom in the lower-scoring languages such as Thai, Arabic, and Serbian. Full per-language and long-input comparisons are in \appref{app:details}.

\subsection{Multilingual NanoBEIR (primary evaluation 2) and auxiliary comparison on HAKARI-Bench}
\label{sec:mnanobeir}

We compare the unified set (\secref{sec:models}; 13 models including BM25) on the second primary evaluation, Multilingual NanoBEIR (14 languages; datasets not by this author, \secref{sec:benchmarks}), and on the task sets of the author-released HAKARI-Bench (\secref{sec:evalbench}; self-citation). All numbers are extracted from HAKARI-Bench's public leaderboard aggregation (\nDCG{}, 0--1). Only the two -pt rows are reference values not on the public leaderboard, from local evaluation under the identical protocol. The columns of Table~\ref{tab:hakari} measure: Overall = the public HAKARI-Bench leaderboard total (the leaderboard's ``micro'' aggregation, i.e., an unweighted mean of task-level \nDCG{} over all 538 deduplicated raw tasks; since the 182 MNanoBEIR cells enter individually, that benchmark holds about 34\% of the weight; a macro aggregation over benchmarks flips some near-tie rankings); MNanoBEIR = the condensed BEIR in 14 languages (general multilingual web retrieval); NanoMMTEB-v2 = a sampled subset of only the Retrieval tasks of MMTEB Multilingual v2 (18 tasks); NanoRTEB = a Nano-ization of the public English part of RTEB \citep{liu2025rteb}, a production-domain retrieval benchmark (14 tasks); NanoCoIR = a Nano-ization of the code-retrieval benchmark CoIR \citep{li2024coir} (10 tasks).

\begin{table}[t]
\caption{Multilingual NanoBEIR (primary evaluation 2) and HAKARI-Bench task sets (\nDCG{}; ascending AP). The MNanoBEIR column is primary evaluation 2; the other columns (Overall, NanoMMTEB-v2, NanoRTEB, NanoCoIR) are auxiliary comparisons on the author-released HAKARI-Bench. The -pt rows are stage-1-only (pretrained) reference values. The released Bekko rows are bold for visibility, not to mark column bests. Long-input task sets appear in \secref{sec:longdoc} (Figure~\ref{fig:longtask}) and \appref{app:details}.}
\label{tab:hakari}
\centering
\small
\setlength{\tabcolsep}{2.2pt}
\begin{tabular}{lrrrrrrr}
\toprule
Model & AP & dims & Overall & MNanoBEIR & NanoMMTEB-v2 & NanoRTEB & NanoCoIR \\
\midrule
BM25 & --- & --- & 0.477 & 0.465 & 0.455 & 0.355 & 0.544 \\
bekko-a8m-pt & 7.7M & 384 & 0.523 & 0.505 & 0.494 & 0.527 & 0.727 \\
bekko-a8m & 7.7M & 384 & \textbf{0.545} & \textbf{0.526} & \textbf{0.502} & \textbf{0.550} & \textbf{0.747} \\
mE5-small & 21.6M & 384 & 0.517 & 0.512 & 0.445 & 0.471 & 0.692 \\
bekko-a25m-pt & 24.9M & 384 & 0.549 & 0.533 & 0.488 & 0.571 & 0.769 \\
bekko-a25m & 24.9M & 384 & \textbf{0.570} & \textbf{0.549} & \textbf{0.494} & \textbf{0.594} & \textbf{0.786} \\
granite-97m-r2 & 28.3M & 384 & 0.525 & 0.505 & 0.531 & 0.567 & 0.780 \\
mE5-base & 86.0M & 768 & 0.539 & 0.531 & 0.469 & 0.514 & 0.700 \\
harrier-270m & 100.3M & 640 & 0.555 & 0.523 & 0.522 & 0.550 & 0.789 \\
embgemma-300m & 106.3M & 768 & 0.597 & 0.577 & 0.517 & 0.670 & 0.847 \\
granite-311m-r2 & 110.3M & 768 & 0.569 & 0.543 & 0.577 & 0.606 & 0.814 \\
gte-m-base & 113.3M & 768 & 0.563 & 0.527 & 0.486 & 0.558 & 0.753 \\
mE5-large & 303.9M & 1024 & 0.565 & 0.560 & 0.484 & 0.556 & 0.747 \\
arctic-l-v2.0 & 311.8M & 1024 & 0.598 & 0.584 & 0.502 & 0.549 & 0.707 \\
bge-m3 & 311.8M & 1024 & 0.577 & 0.557 & 0.485 & 0.536 & 0.692 \\
\bottomrule
\end{tabular}
\end{table}

The highlights are as follows.

\begin{itemize}[leftmargin=*]
\item Multilingual NanoBEIR (primary evaluation 2): the top scores are arctic-l-v2.0 (0.584) and embeddinggemma (0.577). a25m (0.549) is within 0.02 of bge-m3 (0.557) and mE5-large (0.560), which have about 12$\times$ its AP, and beats granite-311m (0.543), mE5-base (0.531), gte-m-base (0.527), and harrier-270m (0.523). a8m (0.526), with 7.7M AP, beats mE5-small (0.512) and granite-97m (0.505) and matches gte-m-base and harrier-270m. Every dense model beats BM25 (0.465).
\item HAKARI Overall (538 raw tasks, micro): the order is arctic (0.598) $\approx$ embeddinggemma (0.597) $>$ bge-m3 (0.577) $>$ a25m (0.570) $\approx$ granite-311m (0.569); a25m sits within 0.01 of bge-m3 and granite-311m, whose AP is 4--12$\times$ its own. a8m at 0.545 beats mE5-small (0.517), granite-97m (0.525), and mE5-base (0.539), within 0.01 of harrier-270m (0.555). Under macro aggregation over benchmarks, near-ties reorder (gte-m-base $\approx$ bge-m3 $\approx$ a25m), so hairline rankings depend on the aggregation scheme.
\item MMTEB-style retrieval (NanoMMTEB-v2) and production-style retrieval (NanoRTEB): on NanoMMTEB-v2 the granite R2 family is strong (311m 0.577 / 97m 0.531); a8m (0.502) ties arctic (0.502), and a25m (0.494) also edges the mE5 family (0.445--0.484). The simple-mean reversal of a8m over a25m stems from a few outlier tasks where a25m drops sharply (three tasks, e.g., 8K-context passkey retrieval); a25m beats a8m on 14 of the 18 tasks and ranks higher in the leaderboard's per-task rank aggregation (Borda). On NanoRTEB, a25m (0.594) is third after embeddinggemma (0.670) and granite-311m (0.606), and a8m (0.550) approaches mE5-large (0.556).
\item Code retrieval (NanoCoIR): after embeddinggemma (0.847), granite-311m (0.814), and harrier-270m (0.789), a25m (0.786) is level with granite-97m (0.780); a8m (0.747) equals mE5-large (0.747).
\end{itemize}

Measuring the stage-1 pretrained models under the same conditions (the -pt rows of Table~\ref{tab:hakari}), a8m-pt $\to$ a8m rises from 0.523 $\to$ 0.545 on Overall, 0.505 $\to$ 0.526 on MNanoBEIR, 0.481 $\to$ 0.545 on NanoMLDR, and 0.656 $\to$ 0.682 on NanoLongEmbed. a25m-pt $\to$ a25m likewise rises 0.549 $\to$ 0.570, 0.533 $\to$ 0.549, 0.530 $\to$ 0.571, and 0.675 $\to$ 0.706. Stage 2 lifts the aggregate scores by about 0.02, with the largest gains in long-document retrieval. On MMTEB, Mean(Task) rises 55.0 $\to$ 56.7 for a8m and 56.7 $\to$ 58.3 for a25m, with the largest improvement in Reranking (42.3 $\to$ 60.6 / 44.4 $\to$ 61.6; Table~\ref{tab:mmteb}). These are not ablations isolating design factors, so we do not attribute the gains causally to the long hard negatives alone.

Summarizing the head-to-head with the closest competitor, granite-97m-r2 (28.3M AP; \secref{sec:efficient}): granite-97m beats a25m on MMTEB Retrieval (60.3 vs 57.5), NanoMMTEB-v2 (0.531 vs 0.494), and MMTEB Clustering (43.6 vs 43.0). Conversely, a25m wins on the MMTEB overall Mean (58.3 vs 51.9), BitextMining (75.4 vs 44.2), STS (73.4 vs 65.6), MNanoBEIR (0.549 vs 0.505), HAKARI Overall (0.570 vs 0.525), and both long-input benchmarks (\secref{sec:longdoc}); even the smaller a8m (7.7M AP) beats granite-97m on MNanoBEIR (0.526) and Overall (0.545) (Tables~\ref{tab:mmteb}--\ref{tab:hakari}). That is, granite-97m is sharply specialized toward MMTEB-style retrieval, whereas Bekko is ahead on the broad retrieval spectrum---multilingual web retrieval, long inputs, code, cross-lingual---and on overall balance, plus the design-side differences of no distillation, released training data, and CPU speed (\secref{sec:speed}: a8m 364 vs granite-97m 125 docs/s).

Because the HAKARI-side numbers rest on an author-released benchmark, interpretation always pairs them with the independent evaluation (\secref{sec:mmteb}). Placing the two side by side: on retrieval-type tasks Bekko rivals or beats much larger-AP models, while on the MMTEB overall mean it is mid-pack---the MMTEB overall view is stricter than what HAKARI Overall suggests, as stated in \secref{sec:mmteb}. The efficiency frontiers of MNanoBEIR and HAKARI Overall versus AP are shown in Figure~\ref{fig:pareto}(c)(d).

\subsection{Long-context and long-document retrieval: results and discussion}
\label{sec:longdoc}

We compare the unified set on two task sets with long inputs. NanoLongEmbed (6 tasks; a Nano-ization of LongEmbed \citep{zhu2024longembed}, including needle-in-a-haystack-style synthetic tasks plus summarization/QA tasks) measures retrieval over long-document corpora; NanoMLDR (13 languages; a Nano-ization of MLDR) measures multilingual retrieval, over a corpus of long documents, of the document containing the target passage (per-task / per-language details in \appref{app:details}).

\begin{figure}[t]
  \centering
  \includegraphics[width=\linewidth]{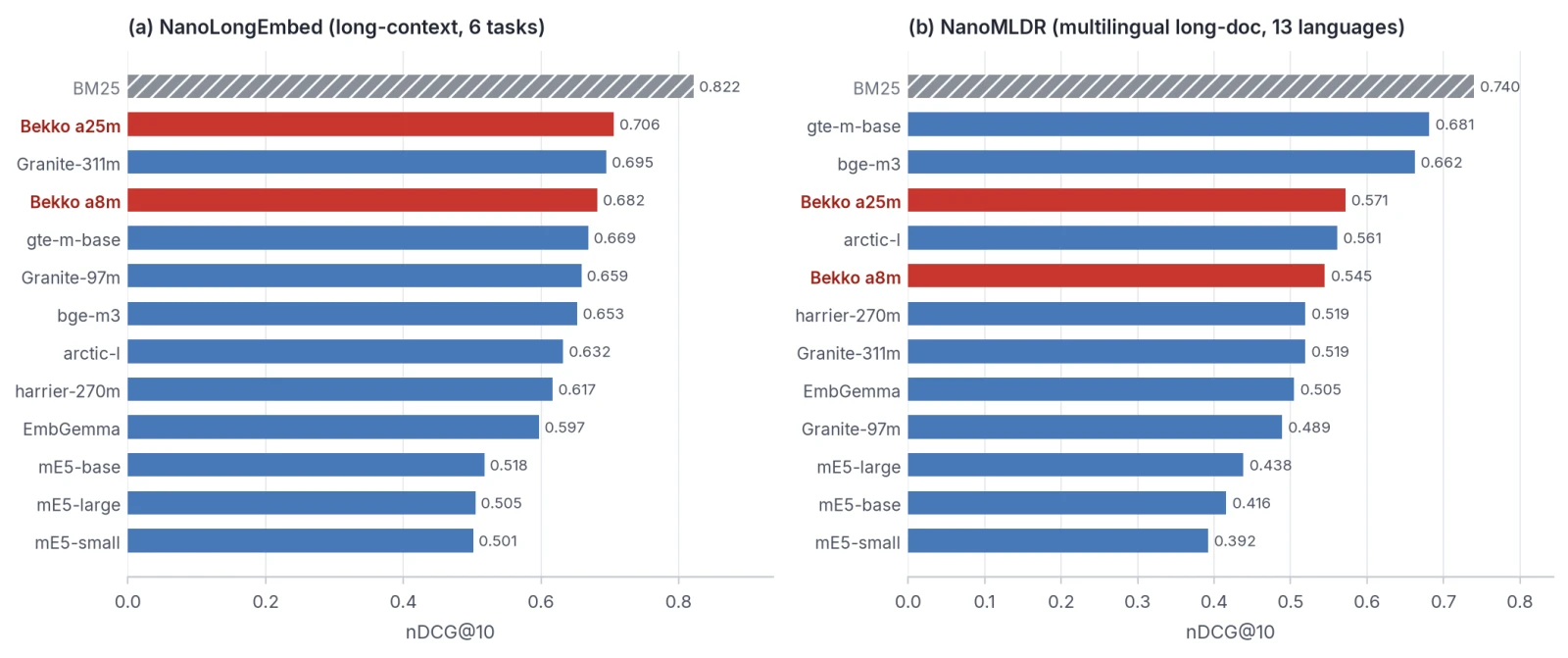}
  \caption{Long-input retrieval (NanoLongEmbed) and long-document retrieval (NanoMLDR). BM25 (hatched) beats every dense model on both benchmarks; among dense models a25m is 1st on NanoLongEmbed and 3rd on NanoMLDR. Full per-task and per-language numbers are in \appref{app:details}.}
  \label{fig:longtask}
\end{figure}

The results (Figure~\ref{fig:longtask}):

\begin{itemize}[leftmargin=*]
\item BM25: in contrast to the short-text-centric MNanoBEIR, where every dense model beat BM25 (\secref{sec:mnanobeir}), BM25 beats all dense models on both long benchmarks (NanoLongEmbed 0.822 / NanoMLDR 0.740). When the task is to hit one passage inside a long document, lexical overlap between query and passage is a strong signal, structurally favoring lexical BM25. The BGE-M3 paper likewise reports BM25 well above dense baselines on MLDR \citep[Table 3]{chen2024bgem3}, indicating that dense long-document retrieval remains an open problem. In production this is territory where hybrid BM25 + dense retrieval pays off, and Bekko slots in cheaply as the dense component.
\item NanoLongEmbed: a25m at 0.706 is 1st among dense models (a8m at 0.682 is 3rd, after granite-311m at 0.695). Per task, a25m scores Nano2WikiMultihopQA 0.873, NanoNarrativeQA 0.480, NanoNeedle 0.688, NanoPasskey 0.752, NanoQMSum 0.474, NanoSummScreenFD 0.969 (\appref{app:details}). The mE5 family, with a 512-token maximum input, is far lower on long-input tasks generally (mE5-large averages 0.505).
\item NanoMLDR: a25m (0.571) is the 3rd dense model after gte-m-base (0.681) and bge-m3 (0.662). Note that MLDR is a benchmark constructed in the bge-m3 paper, and that model's fine-tuning includes long-document data \citep{chen2024bgem3}. Bekko deliberately avoids MLDR-family training data (our long-document hard negatives borrow only MLDR's construction methodology; \secref{sec:data}) and under that condition still beats arctic (0.561), harrier-270m (0.519), the granite R2 family (97m 0.489 / 311m 0.519), embeddinggemma (0.505), and the mE5 family (0.392--0.438). a8m (0.545) beats mE5-large (0.438), which has 40$\times$ its AP, by +0.11.
\end{itemize}

Discussion and hypotheses: Bekko's long-input robustness plausibly draws on both (1) the base mmBERT setting RoPE theta to 160{,}000 for both Global and Local layers, a long-context-oriented configuration (\secref{sec:mmbert}), and (2) stage-2 adaptation to long hard negatives of up to 8192 tokens (11 languages). These factors (RoPE setting, long-document data, task composition, max length) are not disentangled in this paper; separating their contributions is future work (\secref{sec:limitations}).


\section{Analysis: Which Layers Carry Retrieval, and How Far Does the Recipe Go}
\label{sec:analysis}

This section substantiates the design choices. \secref{sec:layers} validates the retained-layer selection underlying the released models (a8m = L4 [0,1,2,18]; a25m = L13 [0--11,18]) against candidate patterns. \secref{sec:recipeablation} validates the main recipe elements with controlled comparisons under the same conditions as the released models' training, and \secref{sec:qat} reports the with/without comparison of quantization-aware training (QAT).

\subsection{Which layers carry retrieval representations: pruning patterns}
\label{sec:layers}

Which layers to keep is decisive for extreme compaction. From mmBERT-small's 22 layers we form candidate retained patterns (5 at L4, 11 at L7, 10 at L13; all candidates and numbers in \appref{app:pruningpatterns}) and apply a probe common to all candidates: after pruning, fine-tune for retrieval on MS MARCO and compare the mean \nDCG{} over NanoBEIR-en (13 tasks; measured values published on the model cards of the released pruned models). The comparison also includes the intermediate L7 (7-layer) size, which is not used in any released model. Main results:

\begin{center}
\footnotesize
\setlength{\tabcolsep}{4pt}
\begin{tabular}{llrp{4.9cm}}
\toprule
Depth & Retained layers & NanoBEIR-en mean & Notes \\
\midrule
22 (no pruning) & all layers & 0.5151 & baseline \\
L13 & [0--11, 18] & 0.4964 & adopted (a25m base); retains 96.4\% of the 22-layer score \\
L13 & [0--11, 21] & 0.4800 & tail changed to final layer \\
L13 & [0--12] (contiguous) & 0.4553 & no deep Global \\
L7 & [0--5, 18] & 0.4693 & adopted (intermediate dev size) \\
L7 & [0--5, 21] & 0.4629 & tail changed to final layer \\
L7 & [0--6] (contiguous) & 0.4291 & no deep Global \\
L4 & [0, 1, 2, 18] & 0.4530 & adopted (a8m base) \\
L4 & [0, 1, 2, 21] & 0.4558 & tail changed to final layer \\
L4 & [0, 1, 2, 3] (contiguous) & 0.3329 & no deep Global \\
L4 & [18, 19, 20, 21] (deep only) & 0.4130 & no early layers \\
\bottomrule
\end{tabular}
\end{center}

Three observations emerge. First, the top candidates at every depth keep ``early layers + a deep Global layer''; configurations discarding the early layers (L4 [18--21] 0.4130, L13 [9--21] 0.4307) degrade sharply. The adopted front-heavy design (early contiguous layers + one deep Global layer) is a simple near-best member of this top group; it is the strict best only at L13 (13C)---at L4, 4C is marginally higher (+0.003), and at L7 the non-contiguous 7H is marginally higher (\appref{app:pruningpatterns}). Second, contiguous configurations with no deep Global layer at all fall far behind ($-$0.04 at L7, $-$0.12 at L4)---within this probe's candidate set, keeping one deep Global layer for long-range dependencies scores consistently high. Third, for the Global layer kept at the tail, the distant deep layer (18) beats the final layer (21), and the gap widens with depth (L13 +0.016, L7 +0.006). At the minimal L4, though, the gap is within noise and reversed ($-$0.003), so ``18 always beats 21'' does not hold. We adopted [0,1,2,18] as a8m's base because, although [0,1,2,21] is marginally higher at L4 alone, the difference is noise-level, and we prioritized design consistency with the G18 configuration that consistently wins at L7 and L13 (\appref{app:pruningpatterns}).

For reference, the scores of the adopted patterns immediately after pruning (before any further training) on NanoMIRACL / NanoFineWeb2IR (non-public development datasets) are 0.318/0.431 for L4 [0,1,2,18], 0.395/0.531 for L7 [0,1,2,3,4,5,18], and 0.498/0.667 for L13 [0--11,18], increasing monotonically with depth. Even models that have merely been pruned thus retain signal useful for retrieval.

The final layer (21) may be specialized to the MLM pretraining output and less general as a retrieval representation. We therefore consider keeping the distant deep Global layer (18) an effective design heuristic, without claiming that retrieval knowledge is causally localized there. The probe is also English-based (MS MARCO / NanoBEIR-en), differing from the released models' multilingual two-stage training. The observation is consistent with the finding that many deep layers are redundant \citep{men2024shortgpt, gromov2024unreasonable} (\secref{sec:efficient}).

\subsection{Validating the training recipe: controlled comparisons in the release lineage}
\label{sec:recipeablation}

Among the main recipe elements, prefix and QAT were checked with controlled comparisons in the release lineage using the full stage-1 data (only the MRL weights use a 20\%-data proxy; \appref{app:ablations}). The prefix comparison spans three conditions including the released ``none''; the QAT comparison holds everything but QAT fixed within a configuration that attaches prefixes (\model{query:~}/\model{passage:~}), sharing all data and MRL settings with the released models except the prefix. The metric is the development evaluation run periodically during training (a composite of several Nano IR evaluations, 0--1), read at the 1-epoch endpoint (final) and at the maximum over training (best). Each setting is a single run; no seed replications were performed (primary data in \appref{app:ablations}).

\begin{itemize}[leftmargin=*]
\item Prefix: for a8m stage 1 we compared ``query + passage (final 0.578 / best 0.581)'', ``none (the released setting; 0.577 / 0.582)'', and ``query only (0.575 / 0.583)'' (\appref{app:ablations}). All differences are within about 0.003. A controlled comparison through stage 2 on MMTEB Multilingual v2 (identical stage-2 data, prefix the only change) also shows a Mean(Task) difference below 0.003 (no-prefix marginally higher): prefix presence makes no substantial difference. As a side note, no-prefix tends to score higher on symmetric tasks such as Clustering (\appref{app:ablations}). Given essentially equal quality, the released models adopt the simpler, easier-to-use no-prefix setting, which does not ask callers to distinguish query from document (\secref{sec:recipe}).
\item MRL weights: in the weight comparison at the release dims (a8m-like configuration, 20\%-data proxy; \appref{app:ablations}), differences across settings are small at 0.001--0.007; sensitivity is low.
\item QAT: the with/without controlled comparison is reported in \secref{sec:qat} (the largest effect among recipe elements).
\item Temperature and learning rate: $\tau$=0.03 and the higher learning rates for smaller models (a8m stage 1 $9.0\times10^{-4}$ $>$ a25m $3.0\times10^{-4}$) are settled values from early development exploration (table in \secref{sec:recipe}). That exploration used configurations different from the released ones, so no controlled comparison is presented.
\end{itemize}

\subsection{Quantization-aware training (QAT): with vs.\ without}
\label{sec:qat}

To build quantization tolerance in from training, we add an auxiliary loss (the QAT loss) that applies the contrastive loss to embeddings quantized to int8 and ubinary in addition to float32 (weights [float32, int8, ubinary] = [1.0, 0.1, 0.1]). The implementation applies the simulated quantization of \citet{jacob2018quantization} to the sentence-embedding space: per-dimension asymmetric quantization onto 256 levels passed through a straight-through estimator (ubinary is binarization to \{0,1\} thresholded at zero).

QAT presence was checked in a controlled comparison within the prefixed (\model{query:~}/\model{passage:~}) configuration with everything else fixed (data and MRL settings identical to the released models except the prefix; the same development evaluation as \secref{sec:recipeablation}; one run per setting; primary data in \appref{app:ablations}). On a8m (L4) stage 1, QAT-on (final 0.578 / best 0.581) clearly beats QAT-off (0.560 / 0.561). On a25m (L13) stage 1 they are nearly equal (0.590 / 0.594 vs 0.587 / 0.596). Further, for L13, carrying each branch through stage 2 on identical FT data (the development-time FT configuration; \appref{app:ablations}) again favors QAT-on (final 0.604) over QAT-off (0.595). In sum, adding QAT does not hurt the reported final values (only the L13 stage-1 best value slightly favors QAT-off, so readings are mixed), and on the minimal L4 the development evaluation is clearly higher with QAT---possibly because the normalization accompanying int8/ubinary quantization acts as a regularizer, though with one run per setting we do not assert this.

The contribution to quantization tolerance itself, by contrast, is not isolated. The released models are nearly lossless under int8/binary quantization (with rescoring) (\secref{sec:vectorcompress}), but whether that is due to QAT or to embeddings having intrinsically information-dense distributions cannot be determined. Also, simulated quantization during training is not guaranteed to match the quantized scoring of production search engines, so quantization tolerance is verified by direct measurement of output vectors (\secref{sec:vectorcompress}).

\section{Compact Model Distribution and Inference Speed}
\label{sec:efficiency}

This section demonstrates on real hardware the consequences of the AP axis (\secref{sec:ap-axis}): compact model distribution (\secref{sec:int8}) and inference speed (\secref{sec:speed}). Bekko ships in two forms: general-purpose PyTorch weights (safetensors), and ONNX / OpenVINO for low-spec and edge environments. In the latter, the static vocabulary embedding table---the bulk of distribution size and load-time memory---is compressed with int8, shrinking the model file (\secref{sec:int8}). Inference is consistently fast from CPUs and edge devices to GPUs thanks to the small AP (\secref{sec:speed}). \secref{sec:weightquant} and \secref{sec:vectorcompress} then separate two operations that are both called ``quantization'' but differ entirely in target and effect: \secref{sec:weightquant} quantizes the model weights (the compute parameters including Transformer layers) to int8, whereas \secref{sec:vectorcompress} quantizes the output vectors produced by inference to int8/binary. As a corollary of all this, the models also run in the browser (\secref{sec:browser}). Table~\ref{tab:deployment} summarizes the practical conclusions of this section.

\begin{table}[t]
\caption{Deployment artifacts and configurations by environment, as measured in this paper (evidence in the respective sections).}
\label{tab:deployment}
\centering
\small
\setlength{\tabcolsep}{4pt}
\begin{tabular}{p{2.9cm}p{4.0cm}p{3.4cm}p{3.3cm}}
\toprule
Environment / use & Configuration & Why & Caveats \\
\midrule
Server / edge CPU & OpenVINO (vocabulary-embedding int8) & fastest CPU backend (\secref{sec:speed}) & optimization is backend-dependent \\
Apple Silicon & PyTorch (MPS) & MPS is fastest on Mac (\secref{sec:speed}) & OpenVINO if CPU-only \\
Browser & ONNX (vocabulary-embedding int8) + Transformers.js & 124 / 190 MiB distributable; runs on WebGPU/WASM (\secref{sec:browser}) & first-download time and peak memory unmeasured \\
GPU & PyTorch (bf16 / fp16) + FlashAttention-2 & highest throughput (\secref{sec:speed}; measured in fp16) & needs CUDA / FlashAttention-2 \\
Vector DB (index compression) & 256 dims + int8 + rescore & near-lossless compression (\secref{sec:vectorcompress}) & fp32 vectors must be kept for rescoring \\
\bottomrule
\end{tabular}
\end{table}

\subsection{Memory: int8 quantization of the static token embedding matrix}
\label{sec:int8}

Bekko's AP is small, but the 256{,}000-vocabulary $\times$ 384-dimension multilingual embedding matrix dominates total parameters and distribution size ($256000\times384\times4$ bytes $\approx$ 375~MiB in fp32). \citet{abdaoui2020load} likewise note that in multilingual models most parameters concentrate in the embedding layer, so compressing the vocabulary side directly reduces distribution size and load-time memory. Because this matrix is a lookup table that contributes nothing to the Transformer's matrix multiplications (lookup, dequantization, and memory-bandwidth costs remain; \secref{sec:ap-axis}), it can be compressed aggressively with little accuracy impact. The ONNX / OpenVINO distributions default to the vocabulary-embedding-int8 build, with fp32 / fp16 builds also shipped for comparison (the general-purpose PyTorch distribution stays safetensors).

\begin{figure}[t]
  \centering
  \includegraphics[width=0.85\linewidth]{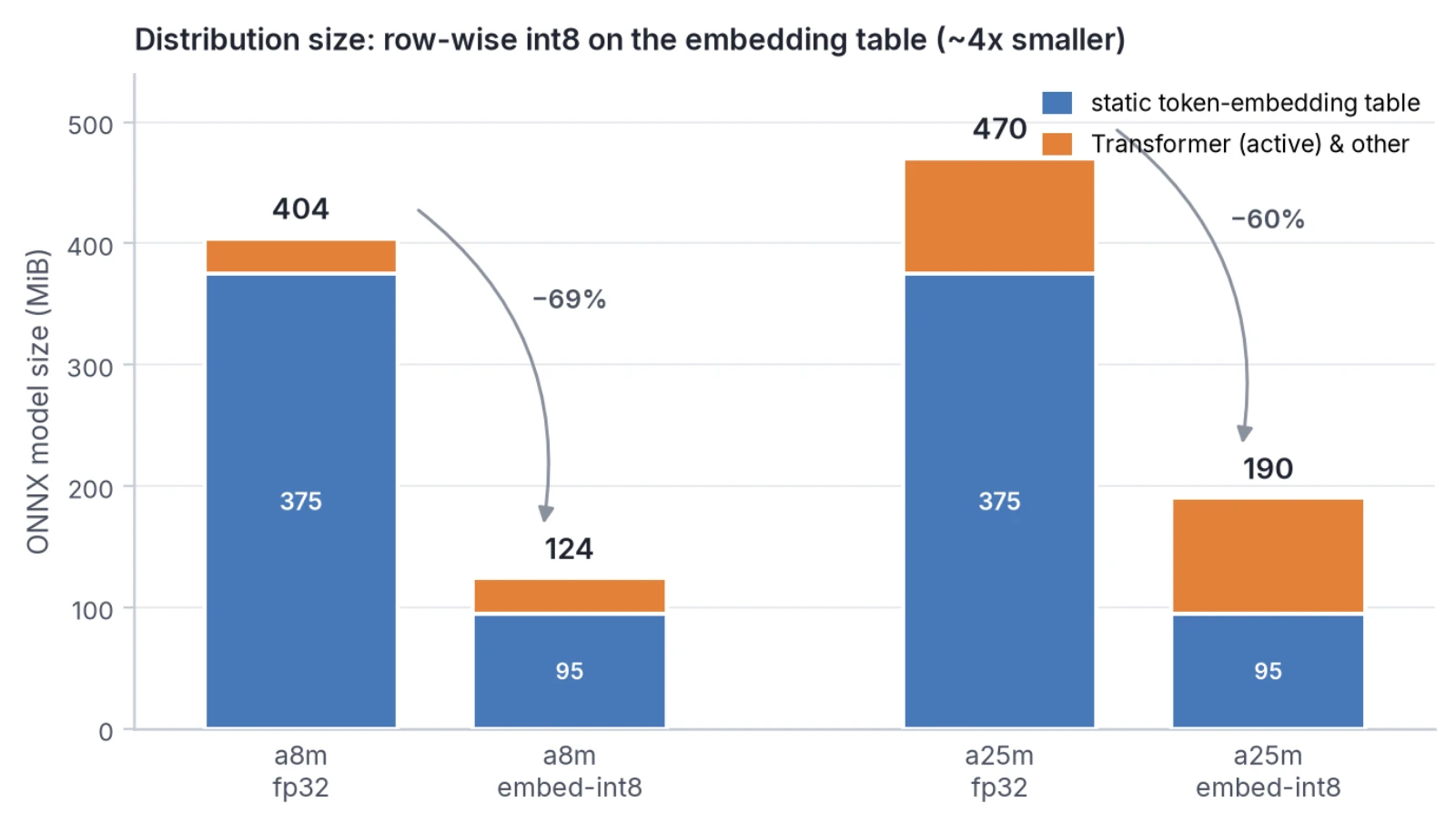}
  \caption{Distribution-size breakdown and the effect of embedding int8. The compression target is the static embedding table, which does not participate in matrix-multiply compute; the AP (Transformer) part is untouched (source: file layout of the public model cards).}
  \label{fig:memory}
\end{figure}

The int8 scheme is row-wise symmetric quantization (distribution-size breakdown and savings in Figure~\ref{fig:memory}). For each token row $\mathbf{w}_r \in \mathbb{R}^{384}$, compute the scale $s_r = \max_j |w_{r,j}| / 127$ and store $\hat{w}_{r,j} = \mathrm{round}(w_{r,j}/s_r) \in [-127,127]$ as int8; reconstruction is $w_{r,j} \approx s_r \cdot \hat{w}_{r,j}$. The stored artifacts are the int8 table ($256000\times384\times1$) plus fp32 row scales ($256000\times4$), cutting the static embedding table by about 74.7\% versus fp32. In the ONNX graph this is implemented as \model{Gather(int8 table)} $\to$ \model{Cast(float)} $\to$ \model{Gather(row\_scale)} $\to$ \model{Mul} $\to$ downstream fp32 Transformer (per-row scales keep quantization error small). We call this build the vocabulary-embedding-int8 version (the default ONNX / OpenVINO distribution).

\begin{center}
\small
\begin{tabular}{lrrrr}
\toprule
Model & fp32 ONNX & Default distribution (vocab int8) & Reduction & tokenizer.json \\
\midrule
a8m & 404.3 MiB & 124.1 MiB & 69.3\% & 32.8 MiB \\
a25m & 470.3 MiB & 190.1 MiB & 59.6\% & 32.8 MiB \\
\bottomrule
\end{tabular}
\end{center}

When this paper says 124 / 190 MiB, it means the ONNX model file alone. Total local footprint / download adds tokenizer.json (32.8 MiB) etc., giving about 157 MiB for a8m and 223 MiB for a25m.

The accuracy cost is tiny: as measured on the released a8m (31 tasks; \secref{sec:weightquant}), this vocabulary-only int8 build differs from the unquantized baseline by $-$0.0001 on average, and the released artifacts are verified for vector consistency against PyTorch (min cosine $\geq$ 0.9994; a8m on x86, Raspberry Pi 5, and Apple Silicon; a25m on x86 and Raspberry Pi 5). Load-time memory is also small, suiting low-resource environments; a8m, whose AP is in the single-digit millions, is lighter still than a25m.

As a further experiment (a preliminary study on the pre-release prefixed lineage at max length 512; \secref{sec:limitations}), pruning the vocabulary itself post hoc from 256{,}000 to 219{,}836 tokens (only tokens observed in FineWiki's 30 languages; vocabulary trimming \citep{ushio2023vocabtrim}) before int8 shrinks the distribution further to 110.7 MiB (a8m) / 176.7 MiB (a25m), with near-zero average loss on the full 231-task Nano set ($-$0.32\% / $-$0.14\%). However, localized degradation appears in low-resource, non-Latin-script languages (e.g., Telugu, Swahili, Bengali NanoMIRACL), so the default distribution keeps the full vocabulary, and the vocabulary-pruned build remains an unreleased experiment (\secref{sec:limitations}).

\subsection{Inference speed: CPU / Raspberry Pi / Apple Silicon / CUDA GPU}
\label{sec:speed}

Our motivation is practicality in GPU-less, low-resource environments, so CPU speed matters most, and we compare it first. Smaller AP means less Transformer compute, and the difference shows directly in CPU throughput (\secref{sec:ap-axis}).

CPU / Apple Silicon / CUDA GPU comparison. The same Natural Questions document encoding (batch 64, max 512 tokens) is measured with the fastest backend per environment---x86 / Raspberry Pi 5: OpenVINO (competitors use their default OpenVINO / int8 artifacts; Bekko the vocabulary-embedding-int8 build); Apple Silicon: PyTorch MPS; CUDA GPU: PyTorch fp16 (both SDPA and FlashAttention-2 shown). The rows are the subset of the unified set (\secref{sec:models}) measured in every environment; mE5-base, gte-m-base, embeddinggemma-300m, harrier-270m, arctic-l-v2.0, and bge-m3 were not measured on x86 / Raspberry Pi under this protocol. CPU / Raspberry Pi speed claims in this section therefore apply to comparisons against the models measured under identical conditions in the table.

\begin{center}
\footnotesize
\setlength{\tabcolsep}{4pt}
\begin{tabular}{lrrrrrr}
\toprule
Model & AP & x86 CPU & Raspberry Pi 5 & Apple M4 Max (MPS) & CUDA (SDPA) & CUDA (FA2) \\
\midrule
bekko-a8m & 7.7M & 364 & 33 & 592 & 4{,}697 & 5{,}561 \\
mE5-small & 21.6M & 226 & 19 & 370 & 3{,}315 & 3{,}746 \\
bekko-a25m & 24.9M & 134 & 10.5 & 351 & 3{,}243 & 4{,}006 \\
granite-97m-r2 & 28.3M & 125 & 10.0 & 286 & 3{,}221 & 3{,}917 \\
granite-311m-r2 & 110.3M & 38 & 2.9 & 106 & 1{,}497 & 2{,}159 \\
mE5-large & 303.9M & 21 & 1.5 & 67 & 1{,}092 & 1{,}318 \\
\bottomrule
\end{tabular}
\end{center}

All units are docs/s (Natural Questions documents encoded per second; higher is faster). x86 CPU = Ryzen 9 7950X (OpenVINO); Raspberry Pi 5 = OpenVINO; Apple M4 Max = PyTorch MPS; CUDA = RTX 5090 (transformers 5.12.1, fp16, SDPA / FlashAttention-2 [FA2]; measured on NQ 100k). Environment details in \appref{app:env}.

\begin{figure}[t]
  \centering
  \includegraphics[width=0.9\linewidth]{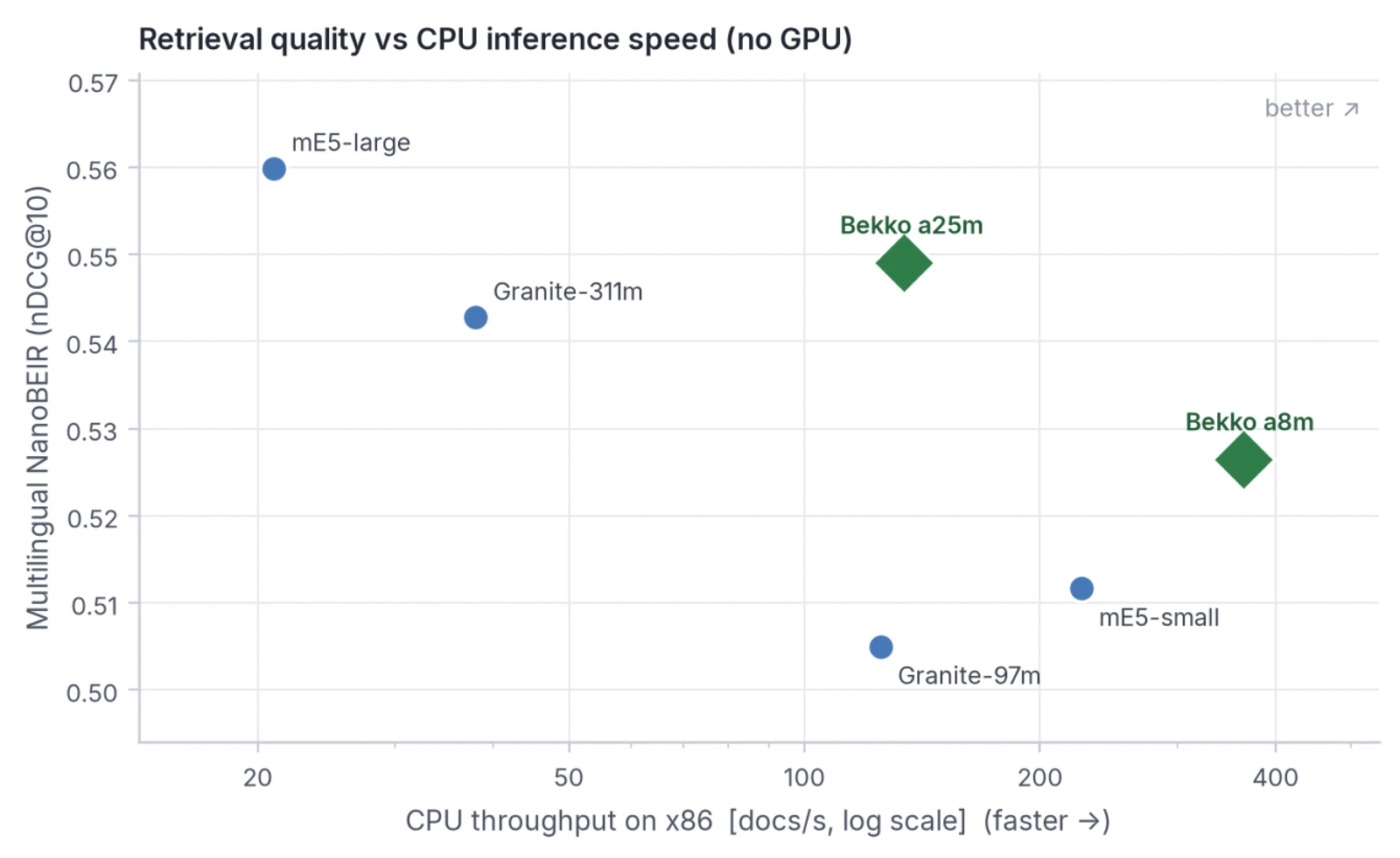}
  \caption{Retrieval quality versus CPU inference speed (GPU-less setting; the models of the unified set with CPU measurements). a8m (364 docs/s) is faster and higher-quality than mE5-small (226); a25m (134) delivers mE5-large-class (21) quality at about 6$\times$ the speed; the two form the quality--speed frontier---the main efficiency claim of this work.}
  \label{fig:cpuspeed}
\end{figure}

Within the measured set, a8m is fastest even on CPU (Figure~\ref{fig:cpuspeed}): on x86 about 1.6$\times$ mE5-small, 2.9$\times$ granite-97m, and 17$\times$ mE5-large; on Raspberry Pi 5 also fastest (about 1.8$\times$ mE5-small, 22$\times$ mE5-large). a25m (x86 134 / Pi 10.5) is roughly level with granite-97m and far above granite-311m and mE5-large, whose AP is 4--12$\times$ larger.

At the same time, the table shows that at similar AP, architecture and inference implementation also matter. mE5-small (AP 21.6M, 226 docs/s), with a conventional BERT (XLM-R-family) architecture, is faster on CPU than the ModernBERT-style a25m (AP 24.9M, 134) and granite-97m (AP 28.3M, 125). ModernBERT's efficiency features---unpadding, FlashAttention variable-length kernels, alternating Global/Local attention---are hardware-oriented designs premised on GPU kernels \citep{warner2024modernbert}; on CPU, where those gains vanish, the simpler conventional BERT is better served by optimized runtimes (OpenVINO, etc.). AP is thus the main driver of FLOPs, but real throughput also depends on architecture, kernels, and backend (consistent with the caveat in \secref{sec:ap-axis}). Even so, a8m pushes its AP down to the single-digit millions, absorbing the architecture difference and remaining the fastest.

On Apple Silicon (the MPS column; environment in \appref{app:env}), a8m is again fastest---about 7.6$\times$ bge-m3 (78 docs/s, outside the table). With CPU only, a8m is also fastest via OpenVINO (297 docs/s). One caveat: with plain PyTorch CPU on the Mac (torch 2.12.1), a8m (76 docs/s) is slower than mE5-small (200). We attribute this to ModernBERT-style kernels being poorly optimized in PyTorch CPU on the Mac (the architecture/implementation factor above); under MPS / OpenVINO the ranking flips back. Results can vary with PyTorch version and build.

CPU backend choice. OpenVINO is the best CPU backend, and within it we default to the vocabulary-embedding-int8 build (\secref{sec:int8}); speed differences among OpenVINO fp32/fp16/vocab-int8 are small, and the int8 build is chosen for its distribution-size and load-memory advantages. Versus PyTorch CPU, OpenVINO speeds up a8m by 2.82$\times$ on x86 (129 $\to$ 364 docs/s) and 1.69$\times$ on Raspberry Pi 5 (19.6 $\to$ 33 docs/s), and a25m by 3.23$\times$ on x86 and 1.74$\times$ on Pi, while cutting model-load RSS by roughly 210--280 MiB. The vocabulary-embedding-int8 ONNX build is used for browser distribution, since browser execution effectively presupposes ONNX Runtime (Transformers.js) (\secref{sec:browser}); on server and edge CPUs, OpenVINO is faster.

Fastest on GPU too. The trend persists on GPU (the CUDA columns). a8m reaches 5{,}561 docs/s with FlashAttention-2, the fastest measured (about 1.5$\times$ mE5-small; about 4.2$\times$ bge-m3 and mE5-large; outside the table, bge-m3 runs 1{,}324 and embeddinggemma-300m 1{,}678 docs/s), and a25m (4{,}006) beats every model except a8m. FlashAttention-2 is +18\% (a8m) / +24\% (a25m) over SDPA.

\subsection{Model weight quantization: vocabulary-only int8 vs.\ full int8}
\label{sec:weightquant}

This section compares two ways of int8-quantizing the model weights (output-vector quantization is treated separately in \secref{sec:vectorcompress}). One is the default-distribution build of \secref{sec:int8}: quantize only the vocabulary embedding table, which dominates distribution size, and leave the Transformer weights in fp32 (below, vocabulary-only int8). The other is post-training quantization of all compute parameters including the Transformer layers (OpenVINO qint8 / ONNX qint8; below, full int8). The former never touches the matrix-multiply compute; the latter quantizes the compute itself---and this difference shows up directly in accuracy, as we demonstrate under identical conditions. We configured the released \model{bekko-embedding-v1-a8m} (the smallest model, the main target of edge distribution) for CPU inference from the same checkpoint as the bf16 baseline and evaluated on NanoBEIR-en + NanoMMTEB-v2 (31 tasks).

\begin{center}
\footnotesize
\setlength{\tabcolsep}{3pt}
\begin{tabular}{lrrr}
\toprule
Configuration & mean \nDCG{} & Abs.\ change & Retention \\
\midrule
Baseline (no quantization; SentenceTransformers / CUDA / bf16) & 0.5448 & --- & --- \\
Vocabulary-only int8 (default distribution; OpenVINO CPU) & 0.5447 & $-$0.0001 & 100.0\% \\
Full int8: OpenVINO qint8 (CPU) & 0.5193 & $-$0.0255 & 95.3\% \\
Full int8: ONNX qint8 AVX512 (CPU) & 0.5113 & $-$0.0335 & 93.8\% \\
\bottomrule
\end{tabular}
\end{center}

Vocabulary-only int8 (the default distribution) is essentially lossless: $-$0.0001 on the 31-task mean (100.0\% retention), at most $-$0.007 on any single task within these 31, and the 8{,}192-token passkey-retrieval task that collapses under full int8 (below) holds at 0.876 $\to$ 0.873. Long-input task sets agree: NanoLongEmbed averages $\pm$0.000 and code retrieval NanoCoIR $-$0.001 (the maximum per-task drop including these is $-$0.008). This is consistent with the artifact vector-consistency checks (min cosine $\geq$ 0.9994 on x86, Raspberry Pi 5, Apple Silicon; \secref{sec:int8}).

Full int8, by contrast, costs 4.7--6.2\% on average, and the damage is uneven: it is largest on the measured long-input task sets (input length, task, language, and backend co-vary, so we do not attribute causality to length alone). About half to 60\% of the 31-task degradation comes from the collapse of the single passkey task (0.876 $\to$ 0.384); excluding it, the remaining 30 tasks average $-$0.010 to $-$0.017. On long-input sets the degradation is stark: $-$8.6\% on long-input retrieval (NanoLongEmbed) and on average $-$23.9\% on long-document retrieval (NanoMLDR, 13 languages; collapse-level in non-Latin-script languages), versus $-$4.8\% on medium-length code retrieval (NanoCoIR). Full int8 thus hits hardest exactly the long-input robustness the released models gained in stage 2 (\secref{sec:longdoc}).

Nor does full int8 buy speed commensurate with that accuracy cost. On a fast x86 CPU (Ryzen 9 7950X, batch 64, max 512 tokens) there is no gain: OpenVINO qint8 (full int8, 397 docs/s) is about 5\% slower than the vocabulary-only-int8 OpenVINO build (420 docs/s)---the CPU speedup comes from the OpenVINO runtime itself, not from int8 Transformer weights. On weak ARM (Raspberry Pi 5, 4$\times$ Cortex-A76), ONNX int8 kernels do speed up 1.6$\times$ over vocabulary-only-int8 ONNX (14.9 $\to$ 23.7 docs/s), but still fall short of vocabulary-only-int8 OpenVINO (28.7 docs/s), while carrying the accuracy and consistency problems above. The speeds in this section were measured for relative comparison across quantization backends on one machine with one script; they are not directly comparable in absolute terms with the table in \secref{sec:speed}, which uses a different measurement stack. Full int8 is also strongly environment-dependent: a25m's OpenVINO qint8 failed the distribution-time vector-consistency gate (cosine 0.98 threshold) and its distribution was withheld; a8m's OpenVINO qint8 produces all-NaN outputs on ARM (Raspberry Pi 5, Apple Silicon); and ONNX qint8 on ARM degrades vector consistency to around cosine 0.77.

In sum, within the measured configurations, on fast and slow CPUs alike, the fastest configuration that preserved accuracy was the vocabulary-only-int8 default distribution (Transformer layers kept in fp32); quantizing all weights including the Transformer offered no benefit commensurate with the accuracy loss.

\subsection{Output vector compression: dimensionality reduction and quantization}
\label{sec:vectorcompress}

Everything up to \secref{sec:weightquant} compressed the model side (weights). This section compresses the output vectors (embeddings) themselves. Output-vector compression directly shrinks vector-database index size and search-time memory/bandwidth, while leaving model weights and inference speed untouched. There are two levers, and they compose:

\begin{itemize}[leftmargin=*]
\item Dimensionality reduction: via MRL (\secref{sec:recipe}), truncate the 384-dimensional output to its first 256 / 128 / 64 dimensions.
\item Output-vector quantization: quantize each fp32 dimension to int8 (1/4 size) or binary (1/32 size).
\end{itemize}

\begin{figure}[t]
  \centering
  \includegraphics[width=\linewidth]{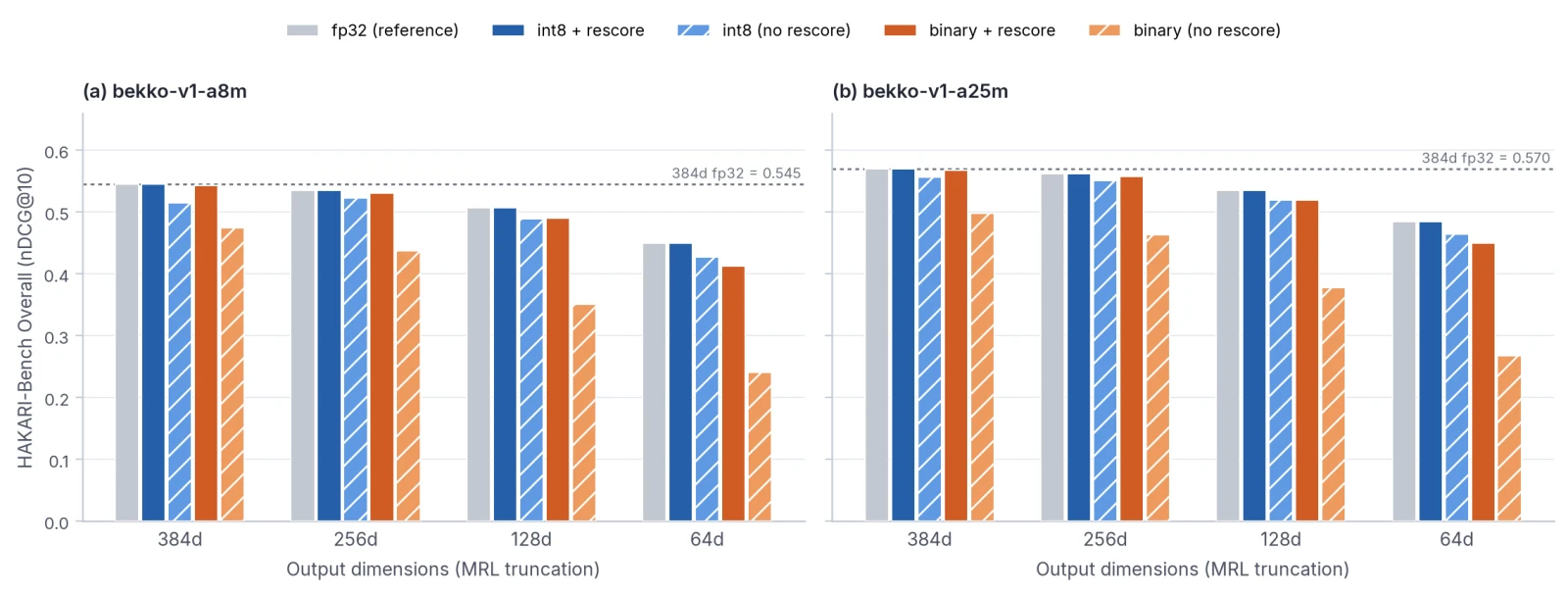}
  \caption{Output-vector compression (dimensionality reduction $\times$ quantization; HAKARI-Bench Overall). Within each dimension group, the int8+rescore bar reaches nearly the fp32 height---the loss converges to the truncation-induced share. Quantization without rescoring loses much more, and binary amplifies the loss at low dimensions.}
  \label{fig:vectorcompress}
\end{figure}

For output-vector quantization we also check how much of the lost retrieval quality rescoring can recover. Rescoring is two-stage retrieval that searches coarsely with quantized vectors and re-scores only the top candidates (top 100 in this evaluation) with fp32 vectors (not a re-search of the whole corpus); it corresponds to the approach of BPR \citep{yamada2021bpr}, which kept accuracy with binary codes plus continuous-vector reranking while cutting memory. We evaluated the full grid for both released models (4 dims $\times$ \{fp32, int8, int8+rescore, binary, binary+rescore\}) on HAKARI-Bench Overall (Figure~\ref{fig:vectorcompress}).

The results reduce to four points (baseline = 384-dim float; a8m 0.5453 / a25m 0.5700; Overall is the same 538-raw-task micro as Table~\ref{tab:hakari}).

\begin{enumerate}[leftmargin=*]
\item Dimensionality reduction is cheap down to 256 dims: $-$1.8\% (a8m) / $-$1.3\% (a25m). At 128 dims it costs $-$7.0\% / $-$6.2\%, and at 64 dims $-$17.5\% / $-$15.0\%.
\item Quantization without rescoring is expensive: even at 384 dims, int8 alone drops $-$5.5\% / $-$2.4\% and binary alone $-$12.9\% / $-$12.6\% (a8m is the more int8-sensitive).
\item With rescoring, near-lossless: int8+rescore costs $-$0.04\% / $-$0.03\%, binary+rescore $-$0.44\% / $-$0.38\%.
\item Combining truncation with quantization, int8+rescore keeps the loss pinned to the truncation share alone (e.g., a25m at 128 dims: float 0.5348 vs int8+rescore 0.5348). Binary, however, amplifies at low dimensions (a25m 64-dim binary+rescore $-$21.1\%, worse than float alone at $-$15.0\%), and unrescored binary at 64 dims ($-$53.0 to $-$55.9\%) is unusable.
\end{enumerate}

As practical guidance, ``256 dims + int8 + rescore'' offers an operating point at which quantization adds almost nothing beyond the truncation cost ($-$1.3 to $-$1.8\%), and binary should be treated as candidate-generation-only, always followed by rescoring. Note that rescoring configurations must keep fp32 (or high-precision) vectors on the side, so the memory savings apply only to the candidate-generation index. Whether the near-perfect recovery under rescoring is due to QAT or to intrinsic properties of the embedding distribution remains unresolved (\secref{sec:qat}). \citet{kisako2026dimquant} also report that the optimal combination of dimensionality reduction and quantization is task-dependent; Bekko provides both MRL truncation and output quantization, preserving deployment-side choice.

\subsection{Running in the browser}
\label{sec:browser}

Despite its huge multilingual tokenizer, Bekko is small---124 MiB (a8m) / 190 MiB (a25m) with ONNX + vocabulary-embedding int8---and runs directly in the browser without a remote server. The implementation is based on Transformers.js: it loads the vocabulary-embedding-int8 \model{onnx/model.onnx}, uses WebGPU where available, and falls back to WASM (CPU). Output dimensions of 384/256/128/64 (via MRL truncation) are supported. The benefits: (1) inputs and documents never leave the device, giving strong privacy; (2) no server or inference-API cost or latency; (3) offline / edge search and RAG become possible. That a multilingual retrieval model loads in a browser at all is a direct consequence of ultra-small AP plus vocabulary-embedding int8.

We measured real browser speed. The environment is Chrome 149 on the same machine as the Apple Silicon measurements of \secref{sec:speed} (Apple M4 Max, macOS 26.5; \appref{app:env}). Encoding 1{,}000 Natural Questions documents (batch 64), compared with native execution on the same machine:

\begin{center}
\small
\begin{tabular}{lrr}
\toprule
Execution mode & a8m & a25m \\
\midrule
Native PyTorch MPS (Apple GPU; \secref{sec:speed}) & 592 & 351 \\
Native OpenVINO (CPU; \secref{sec:speed}) & 297 & 97 \\
Browser WebGPU & 187 & 89 \\
Browser WASM (CPU) & 51 & 16 \\
\bottomrule
\end{tabular}
\end{center}

All units are docs/s (Natural Questions documents encoded per second).

Browser WebGPU runs at about 1/3 (a8m) to 1/4 (a25m) of native MPS speed, and browser WASM at about 1/6 of native OpenVINO CPU. WebGPU is 3.7--5.7$\times$ faster than WASM, and the smaller-AP a8m is about 2.1$\times$ a25m on WebGPU and 3.3$\times$ on WASM---the AP-minimization dividend carries into the browser. This is a simple single-environment, encode-only measurement, not a detailed browser benchmark; its purpose is to demonstrate that a multilingual embedding model runs at practical speed in the browser.

\section{Limitations and Future Work}
\label{sec:limitations}

\begin{enumerate}[leftmargin=*]
\item Skewed language coverage: the training data leans toward English and major languages, and low-resource-language performance is limited. Smaller models likely need their target domains and languages covered directly in the training data, leaving much room to add data for uncovered domains and languages. Strengthening low-resource synthetic data generation is a future direction.
\item Tasks beyond retrieval: Classification sits near the bottom of the unified set, and Clustering / MultilabelClassification stay mid-pack (they improve greatly over the prefixed training configuration, but that is a model-level comparison confounded with other factors; \secref{sec:recipeablation}, \appref{app:ablations}). We view this as a consequence of retrieval-focused data and loss design; improving these task types while preserving retrieval quality is future work.
\item Tokenizer reduction: the 256{,}000-vocabulary embedding matrix still dominates distribution size, leaving room to shrink the vocabulary side. In a local preliminary experiment on an older configuration, post-hoc pruning to observed tokens (\secref{sec:int8}; unreleased) shrank the distribution further with near-zero average loss but left localized degradation in low-resource, non-Latin-script languages. Better retained-vocabulary lists, or retraining with a reduced vocabulary, could push size down while containing that risk.
\item How the base is built: we extracted the base by pruning, but pretraining an ultra-compact configuration from scratch might recover expressiveness lost to pruning. Our work also shows the reach of a distillation-free setup; combining teacher distillation could add further gains.
\item Statistical limits of the ablations: the recipe comparisons of \secref{sec:analysis} (prefix, MRL weights, QAT) each rest on one run per setting, with no seed-variance estimates or significance tests. Differences below 0.01 should be read under that constraint. The QAT-as-regularizer hypothesis is not causally established, and the origin of long-input robustness (\secref{sec:longdoc}; RoPE setting vs.\ long-document data) is unablated.
\item Independence and scale of evaluation data: Multilingual NanoBEIR is a machine-translated direct product of the same 13 English tasks (about 50 queries per task), so its 182 tasks are not independent problems (\secref{sec:evalbench}, \appref{app:details}). The training mixture also contains IR families cognate to evaluations (MS MARCO, mMARCO, MIRACL, HotpotQA, etc.; \secref{sec:data}, \appref{app:stage1}--\ref{app:stage2}), and we have not run a systematic split-level train/eval overlap audit. Anchoring the primary evaluation on official MMTEB mitigates the risk, but a formal overlap audit remains future work.
\item Confounds in the a8m/a25m comparison: the two models share stage-2 data but differ in learning rate, gradient-cache settings, and a25m's checkpoint merge (\secref{sec:recipe}--\secref{sec:singlegpu}), so their difference cannot be read as the effect of AP (capacity) alone (\secref{sec:results}).
\item Scope of the speed evaluation: \secref{sec:speed} measures batch-64 document-encoding throughput; batch-1 query latency (p50/p95), important for interactive search, and length scaling at fixed token counts are not covered. Throughput advantages do not directly imply interactive-latency advantages.
\end{enumerate}

\section{Conclusion}
\label{sec:conclusion}

This work applied Active Parameters (AP)---the chief determinant of inference compute---consistently as the efficiency axis for multilingual embedding models (building on the non-embedding-parameter view of \citealp{kaplan2020scaling} and \citealp{lan2019albert}) and, on that axis, built the ultra-compact multilingual embedding family Bekko Embedding. The method has three pillars. First, without distillation, prune mmBERT-small's 22 layers to 4 (a8m, just under 8M AP) / 13 (a25m, just under 25M AP) with structural layer pruning that preserves the Global-Local rhythm, and train from the pruned model as base. Second, compensate for the small model's knowledge gap with a public corpus of about 1.15 billion multilingual pairs (about 1.11 billion effective; including two complementary LLM-synthesized datasets). Third, run two-stage contrastive training with a loss that integrates the pair-type-dependent masked contrastive loss and MRL, entirely on a single GPU.

In evaluation, on official MMTEB Multilingual v2 Retrieval, a8m (56.2) beats the mE5 family and bge-m3, and a25m (57.5) reaches parity with gte-m-base (57.2), demonstrating retrieval quality that rivals or exceeds that of models with 1--2 orders of magnitude more AP (\secref{sec:mmteb}). The trend is consistent on Multilingual NanoBEIR (14 languages), with strength extending to long-input, long-document, and code retrieval (\secref{sec:mnanobeir}--\secref{sec:longdoc}). On the MMTEB overall mean the models are mid-pack, leaving headroom centered on classification-type tasks. From the design exploration we observed that pruning which keeps early contiguous layers plus a distant deep Global layer---not the final layer---is effective as a design heuristic (\secref{sec:layers}; no claim of causal localization). Controlled recipe comparisons showed that adding QAT costs nothing and, on the minimal a8m, clearly raises the development evaluation (one run per setting; \secref{sec:qat}). On efficiency, among the compared models measured on the same workload and hardware, a8m is the fastest on both CPU and GPU (\secref{sec:speed}), and the ONNX / OpenVINO distributions with the compute-inert vocabulary embedding in row-wise int8 (model files: a8m 124 MiB / a25m 190 MiB) run on a Raspberry Pi 5 and in the browser (\secref{sec:efficiency}). Model weights and training data are openly released; we hope this work serves as a foundation for practical multilingual retrieval in low-resource and on-device settings and for reproducible research on small multilingual embedding models.

\section*{Acknowledgements}

Our training and loss implementations are built on sentence-transformers \citep{reimers2019sbert}. We used its GradCache-style cached contrastive losses, MatryoshkaLoss, and batch-sampler framework, implementing on top of them our masked bidirectional loss, QAT, per-pair-type losses, and token-length-grouped training. The maturity of the sentence-transformers foundation is what made training implementation and evaluation tractable for a single-GPU-scale study. We also thank mmBERT \citep{marone2025mmbert} / ModernBERT \citep{warner2024modernbert} for the base models, and Ruri \citep{tsukagoshi2026ruri} for inspiring the data design.

\bibliographystyle{plainnat}
\bibliography{references}

\clearpage
\appendix

\section{Training and Inference Environments}
\label{app:env}

Training environment:

\begin{center}
\small
\begin{tabular}{p{3.2cm}p{10.6cm}}
\toprule
Item & Value \\
\midrule
GPU (training) & NVIDIA RTX PRO 6000 Blackwell Max-Q $\times$ 1 \\
Training time & Stage 1: a8m $\approx$2.9 days / a25m $\approx$7.7 days. Stage 2: a8m $\approx$1--2 hours / a25m $\approx$3 hours \\
Training software & Python 3.11 / PyTorch 2.8.0 (CUDA 12.9 build; driver CUDA 13.1) / transformers 4.57.6 / sentence-transformers 5.3.x (custom extensions) / flash-attn 2.8.3 / accelerate 1.12.0 / datasets 4.5.0 \\
Framework & custom Trainer built on sentence-transformers \\
\bottomrule
\end{tabular}
\end{center}

Inference-speed measurement environments (primary data for \secref{sec:speed}):

\begin{center}
\small
\begin{tabular}{>{\raggedright\arraybackslash}p{2.3cm}>{\raggedright\arraybackslash}p{4.3cm}>{\raggedright\arraybackslash}p{6.5cm}}
\toprule
Environment & Hardware & Main software \\
\midrule
x86 CPU & AMD Ryzen 9 7950X (16C/32T, AVX-512) & Python 3.11 / PyTorch 2.8.0 / sentence-transformers 5.3.x / ONNX Runtime 1.26.0 / OpenVINO 2026.2.0 \\
Raspberry Pi 5 & ARM Cortex-A76 4 cores (aarch64) & Python 3.11 / PyTorch 2.12.1 / sentence-transformers 5.6.0 / ONNX Runtime 1.27.0 / OpenVINO 2026.2.1 \\
Apple Silicon & M4 Max (macOS 26.5) & Python 3.12 / PyTorch 2.12.1 / sentence-transformers 5.6.0 / ONNX Runtime 1.27.0 / OpenVINO 2026.2.1 \\
GPU (CUDA) & NVIDIA RTX 5090 & Python 3.12 / PyTorch 2.10.0 (CUDA 13.0) / transformers 5.12.1 / sentence-transformers 5.6.0 / FlashAttention 2.8.3 (fp16) \\
Browser & Chrome 149 on Apple M4 Max (macOS 26.5; same machine as the Apple Silicon row) & Transformers.js 4.2.0 (WebGPU / WASM) + ONNX vocabulary-embedding-int8 build \\
\bottomrule
\end{tabular}
\end{center}

The browser measurement is a simple single-environment, encode-only measurement (\secref{sec:browser}).

\section{Stage-1 Data Details (\model{bekko-embedding-v1-unsupervised})}
\label{app:stage1}

We describe the contents of the stage-1 data (the public dataset bekko-embedding-v1-unsupervised) by family. The dataset splits into 2{,}452 subsets (each distributed as a Hugging Face dataset config); each subset's name encodes its source and its pair type (\model{dd}=doc\_doc / \model{qd}=query\_doc / triplet with explicit negatives). Training-time preprocessing applies, to every subset, (1) truncation at 512 tokens and (2) assignment of 8{,}192-row block IDs to avoid within-batch duplicates (the deduplicated blocks of \secref{sec:data}); the parallel-bitext families (from NLLB and CCMatrix) additionally get (3) load-time exclusion of the lowest-scoring language pairs by translation quality (the quality filter of \secref{sec:data}). Every training batch is always drawn from a single subset; subsets are never mixed within a batch (\secref{sec:data}).

By pair type, the subsets comprise 1{,}975 doc\_doc (about 80\%) and 477 query\_doc; 34 subsets across both types are triplets with explicit negatives (\secref{sec:data}; triplets counted within their pair types). doc\_doc subsets are numerous because parallel bitext is split into one subset per language pair (nllb-sampled alone has 1{,}575 subsets); by rows, query\_doc (about 0.66B rows) is larger (\secref{sec:data}).

\begin{center}
\footnotesize
\setlength{\tabcolsep}{4pt}
\begin{tabular}{>{\raggedright\arraybackslash}p{4.3cm}r>{\raggedright\arraybackslash}p{1.55cm}>{\raggedright\arraybackslash}p{5.85cm}}
\toprule
Source & Subsets & Main type & Contents \\
\midrule
\model{nllb-\allowbreak sampled-\allowbreak 500k} & 1{,}575 & doc\_doc & all-pairs bitext from NLLB / CCMatrix (many-to-many, incl.\ low-resource; $\leq$500k rows each) \\
\addlinespace[3pt]
\model{finewiki} & 314 & query\_doc & Wikipedia title$\to$body (314 languages) \\
\addlinespace[3pt]
\model{multilingual-\allowbreak cc-\allowbreak news-\allowbreak ...related-\allowbreak paragraph-\allowbreak pairs} & 134 & doc\_doc & paragraph$\leftrightarrow$paragraph within a news article (134 languages) \\
\addlinespace[3pt]
\model{parallel-\allowbreak sentences-\allowbreak opus-\allowbreak 100} & 99 & doc\_doc & OPUS-100 en$\leftrightarrow$xx parallel sentences \\
\addlinespace[3pt]
\model{nllb-\allowbreak english-\allowbreak bitext-\allowbreak hq} & 95 & doc\_doc & high-quality en$\leftrightarrow$95-language bitext (NLLB / CCMatrix; filtered by bge-reranker-v2-m3 + bge-m3) \\
\addlinespace[3pt]
\model{multilingual\_cc\_news} & 47 & query\_doc & news headline$\to$body (47 languages) \\
\addlinespace[3pt]
\model{parallel-\allowbreak sentences-\allowbreak wikimatrix} & 39 & doc\_doc & WikiMatrix mined bitext \\
\addlinespace[3pt]
\model{parallel-\allowbreak sentences-\allowbreak wikititles} & 1 & doc\_doc & WikiTitles parallel article titles (all language pairs aggregated in one subset) \\
\addlinespace[3pt]
\model{nomic-\allowbreak embed-\allowbreak unsupervised-\allowbreak data} & 25 & qd17 / dd8 & existing weakly supervised pairs (StackExchange/Quora/PAQ/gooaq etc.) \\
\addlinespace[3pt]
\model{FineWeb2-\allowbreak IR} & 21 & qd triplet & query-crafter synthetic queries + BGE-M3 hard negatives (21 languages $\times$ $\approx$3M) \\
\addlinespace[3pt]
\model{swim-\allowbreak ir-\allowbreak \{mono,\allowbreak cross\}-\allowbreak lingual} & 10/17 & query\_doc & SWIM-IR synthetic IR (cross: query and doc languages differ) \\
\addlinespace[3pt]
\model{wikipedia-\allowbreak multilingual-\allowbreak ir-\allowbreak \{pairs,\allowbreak related-\allowbreak lead,\allowbreak related-\allowbreak paragraph\}} & 11 each & qd/dd & title+section$\to$paragraph / lead$\leftrightarrow$related lead / same-article paragraph pairs (11 languages) \\
\addlinespace[3pt]
\model{wikipedia-\allowbreak multilingual-\allowbreak synthetic-\allowbreak ir-\allowbreak query-\allowbreak short\_doc} & 11 & query\_doc & query-crafter synthetic query$\to$short document \\
\addlinespace[3pt]
\model{mmarco-\allowbreak hard-\allowbreak negatives-\allowbreak reranker-\allowbreak filtered} & 10 & qd triplet & multilingual MS MARCO + reranker-filtered hard negatives \\
\addlinespace[3pt]
\model{Nemotron-\allowbreak Post-\allowbreak Training-\allowbreak Dataset-\allowbreak v2} & 9 & query\_doc & instruction$\to$response (chat/code/math/stem + 5 languages) \\
\addlinespace[3pt]
\model{OpenCodeReasoning} & 1 & query\_doc & programming problem$\to$solution pairs \\
\addlinespace[3pt]
\model{lightonai/\allowbreak embeddings-\allowbreak pre-\allowbreak training} & 3 & query\_doc & hermes / beir\_dbpedia / webfaq \\
\addlinespace[3pt]
\model{\{wikipedia-\allowbreak english,\allowbreak fineweb,\allowbreak ccnews,\allowbreak arxiv,\allowbreak pubmed\}-\allowbreak ir-\allowbreak simulated-\allowbreak search-\allowbreak queries} & 1 each & query\_doc & Qwen3.5-35B-A3B synthetic queries (simulating expert search behavior) \\
\addlinespace[3pt]
\model{\{all-\allowbreak nli,\allowbreak wikipedia-\allowbreak synthetic-\allowbreak nli\}} & 1 each & dd triplet & NLI / synthetic hard negatives by factual perturbation \\
\addlinespace[3pt]
\model{miracl} & 1 & qd triplet & MIRACL multilingual \\
\bottomrule
\end{tabular}
\end{center}

Family contents and examples (from the actual data):

\begin{itemize}[leftmargin=*]
\item Parallel bitext (doc\_doc): synonymous sentences in two languages, e.g., en ``When food is gone you are my daily meal'' $\leftrightarrow$ its Japanese translation (nllb-english-bitext-hq), or en$\leftrightarrow$ja subtitle pairs (opus-100). \model{nllb-sampled} is many-to-many with one subset per language pair, such as \model{ace\_Latn-ban\_Latn}. The upstream of the nllb subsets is the NLLB project and CCMatrix mined bitext.
\item doc\_doc (non-bitext, broad positives): paragraph$\leftrightarrow$paragraph within an article (Wikipedia/news), article lead$\leftrightarrow$linked-article lead (related topics). These are not paraphrases but broad same-topic positives (title$\to$body is classified as query\_doc, being asymmetric). This family also includes NLI triplets (all-nli, plus \model{wikipedia-synthetic-nli}, whose synthetic hard negatives subtly alter dates and facts).
\item query\_doc (asymmetric query$\to$document pairs): existing IR datasets (nomic / lighton / swim-ir / miracl / mmarco) plus the two complementary synthetic-query datasets. (1) Fast synthesis (query-crafter-multilingual; \appref{app:querycrafter}) generates several kinds of short queries---keywords, natural-language questions, titles---for each FineWeb2-derived document. Example: ja ``one-on-one lessons in Cebu study abroad'' $\to$ the corresponding web document. (2) High-quality synthesis (Qwen3.5-35B-A3B, NVFP4 quantization) generates context-aware search queries for arXiv, PubMed, CC-News, fineweb-edu, and English Wikipedia documents. Example: arXiv ``sparse PCA algorithm selecting coordinates of largest variance...'' $\to$ the paper's abstract. Both deliberately mix direct and indirect phrasings, fluent questions and keyword strings.
\item Triplets (34 subsets, 1.4\%): with explicit hard negatives (FineWeb2-IR / mmarco / miracl / NLI family).
\end{itemize}

Overall character: language coverage is very broad (finewiki 314 languages, CC-News 134 languages, bitext 1{,}600+ language pairs), and LLM-synthesized queries (query-crafter / Qwen3.5-35B-A3B) hold a large share of the core retrieval signal. Positive ``closeness'' deliberately spans from strict translations to broad same-topic positives---a wide-and-shallow design premised on stage 2 narrowing to high-quality IR data. The quality filters depend on rerankers and embedding models, whose biases may persist. Synthetic and existing IR data center on English and major languages, with parallel bitext and the Wikipedia/news families forming a second layer that covers low-resource languages.

These counts are fixed by measurement of the public data: 1{,}146{,}929{,}152 rows in total ($\approx$1.15B), of which 164{,}478{,}976 rows are our own LLM synthesis and 330{,}678{,}272 rows are parallel bitext over 1{,}622 normalized language pairs. Per-language and per-pair breakdowns can be computed from the public dataset's subsets, which are split by language and language pair.

\section{Stage-2 Data Sources}
\label{app:stage2}

The effective stage-2 (hard-negative fine-tuning) mixture is identical for a8m and a25m: 37 subsets, 1{,}780{,}001 rows in total (measured from the released models' training manifests; the manifests count the ten component datasets of the ``additional'' row below as three grouped subsets, which the table expands by source dataset).

\begin{center}
\footnotesize
\setlength{\tabcolsep}{4pt}
\begin{tabular}{>{\raggedright\arraybackslash}p{2.5cm}>{\raggedright\arraybackslash}p{6.1cm}r>{\raggedright\arraybackslash}p{3.2cm}}
\toprule
Family & Subsets & Rows & Notes \\
\midrule
Ruri v3 FT (10 subsets) & auto-wiki-qa-nemotron 154{,}956 / nli 222{,}429 / jsquad 61{,}659 / quiz-no-mori 37{,}437 / jaquad 31{,}253 / quiz-works 30{,}631 / mkqa 30{,}310 / miracl 25{,}905 / jqara 19{,}515 / mr-tydi 18{,}200 & 632{,}295 & max 512 tokens \\
\addlinespace[3pt]
BGE query\_doc & squad 87{,}599 / hotpotqa 84{,}516 / dureader 80{,}416 / mr-tydi 48{,}729 / miracl 40{,}203 / msmarco 200{,}000 / Chinese mmarco 80{,}000 / other 500 & 621{,}963 & BGE-M3 FT subsets with relatively clear licenses; MS MARCO and Chinese mMARCO capped \\
\addlinespace[3pt]
BGE doc\_doc & BQ 12{,}518 / ATEC 11{,}325 / LCQMC 10{,}000 / PAWSX 9{,}900 & 43{,}743 & as above \\
\addlinespace[3pt]
Own hard negatives (general English) & wikipedia\_hard\_negatives\_english & 250{,}000 & capped sample from the released 500{,}000 rows \\
\addlinespace[3pt]
Own hard negatives (additional query\_doc / doc\_doc / domain) & agnews / gooaq / natural\_questions / trivia\_qa (80{,}000 total); all\_nli / coco\_captions (40{,}000 total); arxiv / codesearch / fineweb / pubmed (80{,}000 total) & 200{,}000 & 20{,}000 per dataset \\
\addlinespace[3pt]
Own hard negatives (long documents, 11 languages) & wikipedia\_hard\_negatives\_long\_docs\_ \{en,\allowbreak fr,\allowbreak es,\allowbreak it,\allowbreak de,\allowbreak pt,\allowbreak ar,\allowbreak ko,\allowbreak ja,\allowbreak ru,\allowbreak zh\} & 32{,}000 & en 12{,}000; 2{,}000 per other language. Max 8192 tokens; queries generated by DeepSeek-V4 (\model{deepseek-v4-flash}) \\
\bottomrule
\end{tabular}
\end{center}

Our mined hard negatives use 482{,}000 rows sampled with the caps above from the public dataset bekko-embedding-v1-hard-negatives (759{,}587 rows, 22 subsets). Batch settings: short-to-medium texts (max 512 tokens) use batch 1152; long documents (max 8192 tokens) use batch 192 (\secref{sec:singlegpu}).

\section{query-crafter-multilingual}
\label{app:querycrafter}

query-crafter-multilingual\footnote{\url{https://huggingface.co/hotchpotch/query-crafter-multilingual}}, used for the fast query synthesis behind FineWeb2-IR and related datasets, is a multilingual query-generation model developed and released by the author (Apache-2.0; 21 languages). Built on Qwen3-1.7B, its purpose is to mass-generate, from a given passage, short search-oriented texts that could retrieve that passage: natural-language questions, keyword queries, FAQs, short summaries, and so on.

Construction has three steps. (1) English uses fineweb-edu and the 20 non-English languages use FineWeb2-HQ as upstream corpora; each passage is clipped to at most 10{,}000 characters and filtered to the 70--700-token range (the ceiling is relaxed only for Greek, whose tokenization inflates). (2) A teacher LLM (DeepSeek-V3 Chat) generates reference outputs for 7 instruction types under an English prompt with explicit groundedness rules---invent no entities, dates, or facts unsupported by the passage; never refer to the subject only by pronoun; return \model{NONE} when evidence is insufficient---yielding an SFT dataset of about 525k rows (released as query-crafter-multilingual-sft-synth-500k\footnote{\url{https://huggingface.co/datasets/hotchpotch/query-crafter-multilingual-sft-synth-500k}}, ODC-By). (3) Qwen3-1.7B / Qwen3-4B are LoRA-fine-tuned on this data (1.7B: rank 16 / alpha 32, lr 1.5e-4, 1 epoch, bf16) and the adapters are merged and released.

The 7 instruction types: query (natural-language question) / alt\_query (paraphrased question) / keywords (3 keywords) / synonym\_keywords (3 keywords with synonym substitution) / title (short title) / faq (FAQ-style question answered by the passage) / summary (short summary of the core facts).

Model selection and quality: on a held-out test (about 6{,}300 rows), the semantic agreement between generated queries and source passages measured by BGE-M3 cosine similarity was 0.7808 for the 1.7B model, 0.7820 for 4B, and 0.7811 for the teacher (DeepSeek-V3)---nearly identical---so the cheaper 1.7B model was released. This cosine evaluation is an offline proxy for semantic agreement, not a retrieval metric (recall / nDCG). Known limitations: generated queries skew toward surface-level phrasing close to the passage, and complex queries requiring inference or background knowledge are hard to produce---this motivates the complementary design that pairs it with the context-aware high-quality synthesis (Qwen3.5-35B-A3B; \secref{sec:data}). Generation carries some noise, e.g., keyword outputs violating the specified word count, or questions whose answers are absent from the source document. These tendencies were identified by manual audits of outputs and removed by rule-based filters (dropping failure-signaling outputs, over-long queries, etc.) before entering training data. For the FineWeb2-IR training data, the five basic types (query, keywords, title, faq, summary) were generated at high ratios and the two paraphrase types at low ratios, and hard negatives for each query were mined from the top neighbors retrieved by BGE-M3.

\section{Pruning Pattern Candidates}
\label{app:pruningpatterns}

All candidates (5 at L4, 11 at L7, 10 at L13) and their scores. Scores use the probe of \secref{sec:layers}: after pruning, fine-tune on MS MARCO for retrieval and compare mean \nDCG{} over NanoBEIR-en (13 tasks). For reference, unpruned mmBERT-small (22 layers) under the same protocol scores 0.5151.

L4 candidates:

\begin{center}
\small
\begin{tabular}{llp{4.5cm}r}
\toprule
Pattern & Retained layers & Design intent & NanoBEIR-en mean \\
\midrule
4A & [0,1,2,3] & contiguous 4 (no deep Global) & 0.3329 \\
4B (adopted) & [0,1,2,18] & contiguous 3 + distant G18 & 0.4530 \\
4C & [0,1,2,21] & contiguous 3 + final G21 & 0.4558 \\
4D & [18,19,20,21] & deep only & 0.4130 \\
4E & [0,19,20,21] & first 1 + deep 3 & 0.4408 \\
\bottomrule
\end{tabular}
\end{center}

L7 candidates:

\begin{center}
\small
\begin{tabular}{llp{4.5cm}r}
\toprule
Pattern & Retained layers & Design intent & NanoBEIR-en mean \\
\midrule
7A & [0,1,2,3,4,5,6] & contiguous 7 (to G6) & 0.4291 \\
7B & [0,1,2,3,4,5,9] & contiguous 6 + G9 & 0.4282 \\
7C & [0,1,2,3,4,5,12] & contiguous 6 + G12 & 0.4204 \\
7D (adopted) & [0,1,2,3,4,5,18] & contiguous 6 + distant G18 & 0.4693 \\
7E & [0,1,2,3,6,7,8] & first 4 + middle block & 0.4236 \\
7F & [0,1,2,3,6,7,12] & first 4 + middle + G12 & 0.4149 \\
7G & [0,1,2,3,6,8,12] & first 4 + strided + G12 & 0.4171 \\
7H & [0,1,2,3,6,8,18] & first 4 + strided + G18 & 0.4722 \\
7I & [0,1,2,3,4,5,21] & contiguous 6 + final G21 & 0.4629 \\
7J & [15,16,17,18,19,20,21] & deep only & 0.4299 \\
7K & [0,16,17,18,19,20,21] & first 1 + deep 6 & 0.4589 \\
\bottomrule
\end{tabular}
\end{center}

L13 candidates:

\begin{center}
\small
\begin{tabular}{llp{4.5cm}r}
\toprule
Pattern & Retained layers & Design intent & NanoBEIR-en mean \\
\midrule
13A & [0--12] & contiguous 13 (to G12) & 0.4553 \\
13B & [0--11, 15] & contiguous 12 + G15 & 0.4576 \\
13C (adopted) & [0--11, 18] & contiguous 12 + distant G18 & 0.4964 \\
13D & [0--11, 21] & contiguous 12 + final G21 & 0.4800 \\
13E & [0--10, 12, 18] & contiguous 11 + G12 + G18 & 0.4904 \\
13F & [0--9, 12, 15, 18] & contiguous 10 + G12/15/18 & 0.4791 \\
13G & [0--8, 10, 12, 15, 18] & first 9 + strided & 0.4877 \\
13H & [0--7, 9, 12, 15, 18, 21] & first 8 + strided + G21 & 0.4810 \\
13I & [9--21] & deep only & 0.4307 \\
13J & [0, 10--21] & first 1 + deep 12 & 0.4806 \\
\bottomrule
\end{tabular}
\end{center}

The adopted patterns are not always the strict best at each depth: at L7 the non-contiguous 7H (0.4722) slightly beats 7D (0.4693), but we adopted 7D for the simplicity of contiguous layers. At L4, 4C (G21) beats 4B (G18) by a noise-level +0.003, but we adopted 4B for the G18 advantage consistent at L7 and L13 and for design uniformity (\secref{sec:layers}).

Also, the probe's 13 tasks include NanoMSMARCO, in-domain with the MS MARCO fine-tuning. As a sensitivity check we recomputed rankings over the 12 tasks excluding it: the rankings are unchanged---4C 0.4530 $>$ 4B 0.4486, and 13C (0.4950) is still the clear best---while 7H and 7D become a near tie at 0.4697 $\approx$ 0.4698 (rank flip). The L13 conclusion is thus stable, while the hairline L4/L7 rankings are sensitive to the in-domain component.

\section{Ablation Details}
\label{app:ablations}

F.1 (prefix) compares three conditions including the released ``none''. F.2 (QAT on/off) is a controlled comparison within the prefixed (\model{query:~}/\model{passage:~}) configuration, sharing the stage-1 data and MRL settings with the released models except the prefix. The metric is the development evaluation during training (a composite of several Nano IR evaluations, 0--1); final = the 1-epoch endpoint, best = the maximum over training. One run per setting. F.1--F.2 are the primary data for \secref{sec:recipeablation}--\secref{sec:qat}; F.3 for the MRL weights (\secref{sec:recipe}).

\subsection{Prefix (a8m stage 1, full data)}
\label{app:prefix}

\begin{center}
\small
\begin{tabular}{lrr}
\toprule
Setting & final & best \\
\midrule
query + passage & 0.5778 & 0.5807 \\
none (released setting) & 0.5768 & 0.5820 \\
query only & 0.5752 & 0.5826 \\
\bottomrule
\end{tabular}
\end{center}

The effect of the prefix through stage 2 was also checked on MMTEB Multilingual v2 (\secref{sec:recipeablation}). That comparison uses an unreleased control model trained with prefixes (query + passage) solely for this comparison, with stage-2 data and settings matched. The Mean(Task) difference is below 0.003 (no-prefix marginally higher), and when each configuration is added on its own to the official 2026-06-28 snapshot (170 models), both land at the same overall rank. By task type, relative to the prefixed configuration, no-prefix keeps Retrieval within one point while trending higher by +6 to +11 points on Clustering and about +6 on MultilabelClassification (this is a model-level reference comparison that includes factors other than the prefix). The adoption rationale: prefix presence makes no substantial difference, and no-prefix is simpler and easier to use.

\subsection{QAT on/off (all else equal; MRL included in both)}
\label{app:qat}

\begin{center}
\footnotesize
\begin{tabular}{lll}
\toprule
Comparison & QAT on: final / best & QAT off: final / best \\
\midrule
a8m stage 1 (L4) & 0.5778 / 0.5807 & 0.5605 / 0.5608 \\
a25m stage 1 (L13) & 0.5900 / 0.5937 & 0.5870 / 0.5955 \\
L13: through stage 2 (identical dev-time FT data) & 0.6043 / 0.6054 & 0.5954 / 0.5992 \\
\bottomrule
\end{tabular}
\end{center}

The FT data used in the ``through stage 2'' comparison is a development-time configuration, different from the released FT recipe (\appref{app:stage2}); the same data was used for both branches.

\subsection{MRL weight comparison (a8m-like configuration, 20\%-data proxy)}
\label{app:mrl}

Comparison of MRL weights at the release dims [384,256,128,64]. This is a proxy ablation with an L4 configuration mimicking a8m stage 1, a 20\% sample of the stage-1 data, and no QAT; final checkpoints are evaluated on the full HAKARI-Bench dense suite (551 tasks, \nDCG{}). 551 is the HAKARI-Bench dense task count at evaluation time (as in that paper), a different snapshot from the public leaderboard pin (538 tasks) used in the release comparison (\secref{sec:mnanobeir})---this table serves only for relative comparison within the proxy and is not directly comparable with \secref{sec:mnanobeir} numbers.

\begin{center}
\footnotesize
\begin{tabular}{lrrr}
\toprule
weights (w384/256/128/64) & HAKARI all (551) & MNanoBEIR (182) & NanoMMTEB-v2 (18) \\
\midrule
1.0 / 0.6 / 0.3 / 0.15 & 0.4968 & 0.4901 & 0.4602 \\
1.0 / 0.5 / 0.2 / 0.1 & 0.4967 & 0.4903 & 0.4622 \\
1.0 / 0.3 / 0.15 / 0.1 (released) & 0.4955 & 0.4890 & 0.4555 \\
\bottomrule
\end{tabular}
\end{center}

Differences across settings are small (0.001--0.007); weight sensitivity is low. The released setting is last of the three on all three metrics, but by slim margins of 0.001--0.007, with mildly stronger low-dimension weights slightly better (a proxy: read trends, not absolute values).

\section{Unified-Set Details: MNanoBEIR per Language and Long-Input Tasks}
\label{app:details}

Detailed scores of the unified comparison set (\secref{sec:models}) underlying the summaries of \secref{sec:results}. Per-task-type MMTEB scores are in Table~\ref{tab:mmteb}. MNanoBEIR and the long-input sets are extracted from HAKARI-Bench's public leaderboard aggregation (standard configuration without quantization or dimension truncation; Bekko: a8m rev \model{953408de}, a25m rev \model{8acf4b74}, no prefix).

\subsection{MNanoBEIR per language}
\label{app:perlang}

(\nDCG{}, 0--1; 14 languages, released models, standard configuration. Bold = the bekko-a8m / bekko-a25m rows, for visibility, not column bests.)

Each \model{NanoBEIR-\{lang\}} in MNanoBEIR derives from community machine translations of NanoBEIR (Zeta Alpha) (LightOn 8 languages / Liquid AI ja, ko / Serbian-AI-Society sr / sionic-ai th, vi), none by this author (\secref{sec:benchmarks}); as datasets they are therefore more independent than the author-released HAKARI-Bench (self-citation). English is sentence-transformers/NanoBEIR-en (an aggregate repack of the 13 original zeta-alpha-ai sets); the 13 translated languages are the NanoBEIR-\{lang\} of the respective orgs; all 14 languages share the same 13 tasks, a complete 14$\times$13 = 182-task direct product, so the mean over languages equals the mean over all 182 tasks.

\begin{center}
\footnotesize
\setlength{\tabcolsep}{5pt}
\begin{tabular}{lrrrrrrrrrr}
\toprule
Model & AP(M) & dims & avg & en & fr & de & it & es & pt & vi \\
\midrule
BM25 & --- & --- & 0.465 & 0.572 & 0.514 & 0.450 & 0.486 & 0.506 & 0.485 & 0.462 \\
bekko-a8m & 7.7 & 384 & \textbf{0.526} & \textbf{0.603} & \textbf{0.538} & \textbf{0.538} & \textbf{0.536} & \textbf{0.552} & \textbf{0.542} & \textbf{0.537} \\
mE5-small & 21.6 & 384 & 0.512 & 0.584 & 0.510 & 0.539 & 0.537 & 0.534 & 0.482 & 0.523 \\
bekko-a25m & 24.9 & 384 & \textbf{0.549} & \textbf{0.617} & \textbf{0.549} & \textbf{0.558} & \textbf{0.563} & \textbf{0.560} & \textbf{0.558} & \textbf{0.556} \\
granite-97m-r2 & 28.3 & 384 & 0.505 & 0.592 & 0.535 & 0.535 & 0.522 & 0.532 & 0.522 & 0.517 \\
mE5-base & 86.0 & 768 & 0.531 & 0.603 & 0.541 & 0.542 & 0.545 & 0.537 & 0.524 & 0.540 \\
harrier-270m & 100.3 & 640 & 0.523 & 0.598 & 0.553 & 0.531 & 0.532 & 0.532 & 0.529 & 0.529 \\
embgemma-300m & 106.3 & 768 & 0.577 & 0.658 & 0.600 & 0.595 & 0.592 & 0.596 & 0.599 & 0.580 \\
granite-311m-r2 & 110.3 & 768 & 0.543 & 0.613 & 0.562 & 0.554 & 0.555 & 0.560 & 0.561 & 0.548 \\
gte-m-base & 113.3 & 768 & 0.527 & 0.625 & 0.545 & 0.524 & 0.531 & 0.540 & 0.537 & 0.540 \\
mE5-large & 303.9 & 1024 & 0.560 & 0.628 & 0.556 & 0.578 & 0.565 & 0.562 & 0.539 & 0.570 \\
arctic-l-v2.0 & 311.8 & 1024 & 0.584 & 0.652 & 0.590 & 0.603 & 0.596 & 0.590 & 0.589 & 0.583 \\
bge-m3 & 311.8 & 1024 & 0.557 & 0.607 & 0.557 & 0.564 & 0.566 & 0.561 & 0.552 & 0.570 \\
\bottomrule
\end{tabular}

\medskip

\begin{tabular}{lrrrrrrr}
\toprule
Model & sr & sv & no & ja & ko & ar & th \\
\midrule
BM25 & 0.396 & 0.422 & 0.423 & 0.468 & 0.450 & 0.436 & 0.437 \\
bekko-a8m & \textbf{0.484} & \textbf{0.520} & \textbf{0.511} & \textbf{0.537} & \textbf{0.506} & \textbf{0.498} & \textbf{0.468} \\
mE5-small & 0.438 & 0.513 & 0.499 & 0.523 & 0.506 & 0.464 & 0.512 \\
bekko-a25m & \textbf{0.529} & \textbf{0.541} & \textbf{0.533} & \textbf{0.550} & \textbf{0.530} & \textbf{0.523} & \textbf{0.520} \\
granite-97m-r2 & 0.441 & 0.504 & 0.484 & 0.497 & 0.495 & 0.461 & 0.434 \\
mE5-base & 0.482 & 0.536 & 0.538 & 0.522 & 0.517 & 0.491 & 0.520 \\
harrier-270m & 0.504 & 0.512 & 0.502 & 0.507 & 0.495 & 0.489 & 0.507 \\
embgemma-300m & 0.514 & 0.544 & 0.542 & 0.588 & 0.565 & 0.552 & 0.557 \\
granite-311m-r2 & 0.508 & 0.543 & 0.522 & 0.551 & 0.531 & 0.499 & 0.492 \\
gte-m-base & 0.508 & 0.529 & 0.517 & 0.513 & 0.495 & 0.477 & 0.491 \\
mE5-large & 0.522 & 0.559 & 0.556 & 0.557 & 0.557 & 0.522 & 0.567 \\
arctic-l-v2.0 & 0.567 & 0.585 & 0.586 & 0.566 & 0.571 & 0.536 & 0.567 \\
bge-m3 & 0.555 & 0.564 & 0.548 & 0.559 & 0.546 & 0.511 & 0.545 \\
\bottomrule
\end{tabular}
\end{center}

\begin{figure}[t]
  \centering
  \includegraphics[width=\linewidth]{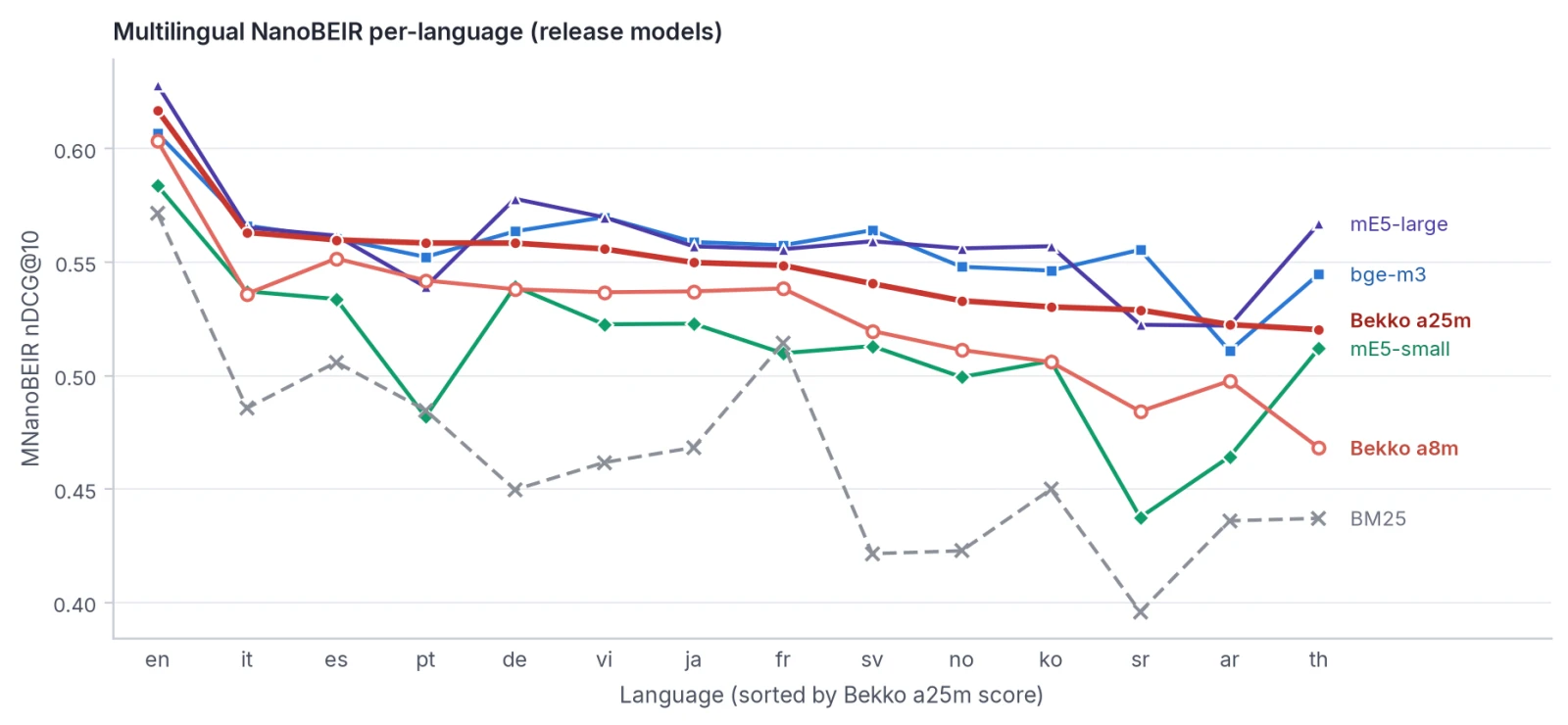}
  \caption{MNanoBEIR per-language scores (6 selected models; all models in the table above). a25m varies relatively little across languages, running broadly alongside bge-m3 and mE5-large, whose AP is about 12$\times$ (max gap to bge-m3: 0.026).}
  \label{fig:perlang}
\end{figure}

Highlights: (1) a25m beats mE5-small, BM25, and granite-97m in all 14 languages, and beats bge-m3 in en, pt, and ar. (2) Its gap to bge-m3 is at most 0.026 (sr); a25m varies relatively little across languages. (3) a8m, at 7.7M AP, beats BM25 in every language and is at or above mE5-small except in de, it, and th (ko ties at three-decimal rounding). (4) The bottom three languages for both Bekko models are th, ar, and sr (Serbian), with ko close behind---mostly non-Latin scripts, but Serbian joins them. The weakest language differs by model, but for Bekko these lower-scoring languages are where further improvement is needed (\secref{sec:limitations}).

\subsection{Long-input task details: NanoMLDR per language, NanoLongEmbed per task}
\label{app:longtask}

(\nDCG{}, 0--1. Bold = the bekko-a8m / bekko-a25m rows, for visibility, not column bests.)

NanoMLDR per language (13 languages):

\begin{center}
\footnotesize
\setlength{\tabcolsep}{5pt}
\begin{tabular}{lrrrrrrrrr}
\toprule
Model & AP(M) & dims & avg & en & de & es & fr & it & pt \\
\midrule
BM25 & --- & --- & 0.740 & 0.725 & 0.714 & 0.944 & 0.912 & 0.888 & 0.950 \\
bekko-a8m & 7.7 & 384 & \textbf{0.545} & \textbf{0.480} & \textbf{0.487} & \textbf{0.781} & \textbf{0.806} & \textbf{0.689} & \textbf{0.806} \\
mE5-small & 21.6 & 384 & 0.392 & 0.395 & 0.276 & 0.540 & 0.567 & 0.558 & 0.608 \\
bekko-a25m & 24.9 & 384 & \textbf{0.571} & \textbf{0.517} & \textbf{0.516} & \textbf{0.818} & \textbf{0.826} & \textbf{0.719} & \textbf{0.840} \\
granite-97m-r2 & 28.3 & 384 & 0.489 & 0.428 & 0.392 & 0.732 & 0.752 & 0.663 & 0.740 \\
mE5-base & 86.0 & 768 & 0.416 & 0.413 & 0.333 & 0.583 & 0.635 & 0.540 & 0.657 \\
harrier-270m & 100.3 & 640 & 0.519 & 0.457 & 0.419 & 0.781 & 0.772 & 0.684 & 0.766 \\
embgemma-300m & 106.3 & 768 & 0.505 & 0.448 & 0.463 & 0.707 & 0.746 & 0.643 & 0.737 \\
granite-311m-r2 & 110.3 & 768 & 0.519 & 0.457 & 0.440 & 0.733 & 0.759 & 0.661 & 0.782 \\
gte-m-base & 113.3 & 768 & 0.681 & 0.548 & 0.600 & 0.908 & 0.890 & 0.842 & 0.906 \\
mE5-large & 303.9 & 1024 & 0.438 & 0.420 & 0.335 & 0.613 & 0.625 & 0.590 & 0.672 \\
arctic-l-v2.0 & 311.8 & 1024 & 0.561 & 0.504 & 0.499 & 0.792 & 0.761 & 0.700 & 0.807 \\
bge-m3 & 311.8 & 1024 & 0.662 & 0.511 & 0.592 & 0.891 & 0.880 & 0.814 & 0.883 \\
\bottomrule
\end{tabular}

\medskip

\begin{tabular}{lrrrrrrr}
\toprule
Model & ru & ar & hi & ja & ko & th & zh \\
\midrule
BM25 & 0.866 & 0.760 & 0.318 & 0.759 & 0.687 & 0.387 & 0.703 \\
bekko-a8m & \textbf{0.653} & \textbf{0.485} & \textbf{0.296} & \textbf{0.555} & \textbf{0.412} & \textbf{0.251} & \textbf{0.382} \\
mE5-small & 0.478 & 0.357 & 0.241 & 0.368 & 0.265 & 0.186 & 0.258 \\
bekko-a25m & \textbf{0.658} & \textbf{0.492} & \textbf{0.348} & \textbf{0.565} & \textbf{0.429} & \textbf{0.319} & \textbf{0.380} \\
granite-97m-r2 & 0.550 & 0.422 & 0.281 & 0.426 & 0.362 & 0.228 & 0.377 \\
mE5-base & 0.490 & 0.392 & 0.230 & 0.385 & 0.292 & 0.183 & 0.272 \\
harrier-270m & 0.598 & 0.443 & 0.319 & 0.502 & 0.409 & 0.262 & 0.341 \\
embgemma-300m & 0.587 & 0.415 & 0.301 & 0.519 & 0.357 & 0.288 & 0.350 \\
granite-311m-r2 & 0.585 & 0.441 & 0.311 & 0.495 & 0.399 & 0.298 & 0.389 \\
gte-m-base & 0.775 & 0.660 & 0.531 & 0.714 & 0.640 & 0.413 & 0.430 \\
mE5-large & 0.504 & 0.417 & 0.268 & 0.402 & 0.337 & 0.236 & 0.277 \\
arctic-l-v2.0 & 0.655 & 0.455 & 0.340 & 0.586 & 0.453 & 0.358 & 0.382 \\
bge-m3 & 0.746 & 0.602 & 0.497 & 0.698 & 0.586 & 0.480 & 0.428 \\
\bottomrule
\end{tabular}
\end{center}

NanoLongEmbed per task (6 tasks):

\begin{center}
\scriptsize
\setlength{\tabcolsep}{3pt}
\begin{tabular}{lrrrrrrrrr}
\toprule
Model & AP(M) & dims & avg & 2WikiMultihopQA & NarrativeQA & Needle & Passkey & QMSum & SummScreenFD \\
\midrule
BM25 & --- & --- & 0.822 & 0.950 & 0.762 & 0.721 & 0.772 & 0.744 & 0.981 \\
bekko-a8m & 7.7 & 384 & \textbf{0.682} & \textbf{0.802} & \textbf{0.441} & \textbf{0.692} & \textbf{0.769} & \textbf{0.420} & \textbf{0.968} \\
mE5-small & 21.6 & 384 & 0.501 & 0.588 & 0.218 & 0.636 & 0.710 & 0.212 & 0.645 \\
bekko-a25m & 24.9 & 384 & \textbf{0.706} & \textbf{0.873} & \textbf{0.480} & \textbf{0.688} & \textbf{0.752} & \textbf{0.474} & \textbf{0.969} \\
granite-97m-r2 & 28.3 & 384 & 0.659 & 0.823 & 0.456 & 0.629 & 0.720 & 0.416 & 0.908 \\
mE5-base & 86.0 & 768 & 0.518 & 0.575 & 0.229 & 0.633 & 0.710 & 0.252 & 0.708 \\
harrier-270m & 100.3 & 640 & 0.617 & 0.840 & 0.334 & 0.599 & 0.640 & 0.370 & 0.919 \\
embgemma-300m & 106.3 & 768 & 0.597 & 0.731 & 0.286 & 0.558 & 0.733 & 0.363 & 0.913 \\
granite-311m-r2 & 110.3 & 768 & 0.695 & 0.872 & 0.521 & 0.612 & 0.726 & 0.473 & 0.965 \\
gte-m-base & 113.3 & 768 & 0.669 & 0.864 & 0.531 & 0.530 & 0.711 & 0.414 & 0.962 \\
mE5-large & 303.9 & 1024 & 0.505 & 0.586 & 0.253 & 0.545 & 0.675 & 0.232 & 0.737 \\
arctic-l-v2.0 & 311.8 & 1024 & 0.632 & 0.770 & 0.416 & 0.518 & 0.734 & 0.392 & 0.965 \\
bge-m3 & 311.8 & 1024 & 0.653 & 0.816 & 0.485 & 0.578 & 0.739 & 0.356 & 0.943 \\
\bottomrule
\end{tabular}
\end{center}

Highlights: (1) BM25 tops nearly every column of both benchmarks---most strikingly the European languages of NanoMLDR (es 0.944 / pt 0.950) and 2WikiMultihopQA / SummScreenFD of NanoLongEmbed---though in NanoMLDR hi (BM25 0.318) and th (0.387) some dense models overtake it. (2) Among dense models, NanoMLDR orders gte-m-base $>$ bge-m3 $>$ a25m (bge-m3's training includes long-document data; \citealp{chen2024bgem3}). On NanoLongEmbed a25m leads the dense average, while per task a8m tops dense models on Needle / Passkey and a25m on 2WikiMultihopQA / QMSum / SummScreenFD, with gte-m-base ahead on NarrativeQA. (3) The mE5 family (max length 512) is far lower on long-input tasks across the board---long-context support is where the architecture generation gap shows most.

\end{document}